\documentclass[final,5p,times,twocolumn]{elsarticle}

\usepackage[T1]{fontenc}
\usepackage[utf8]{inputenc}

\usepackage{amsmath}
\usepackage{amsfonts}
\usepackage{amssymb}
\usepackage{amsxtra}
\usepackage{array}
\usepackage{color}
\usepackage{dcolumn}
\usepackage{graphicx}
\usepackage{hepunits}
\usepackage{xspace}
\usepackage{CJKutf8}

\definecolor{purple}{rgb}{0.5,0,0.5}
\definecolor{blue}{rgb}{0.0,0,0.9}
\definecolor{prdblue}{rgb}{0.133,0.118,0.498}
\usepackage[colorlinks=true, pdfstartview=FitV, linkcolor=prdblue, citecolor= prdblue, urlcolor=prdblue]{hyperref}

\usepackage[mathscr,scaled=1.15]{urwchancal}
\DeclareFontFamily{OT1}{pzc}{}
\DeclareFontShape{OT1}{pzc}{m}{it}%
{<-> s * [1.15] pzcmi7t}{}
\DeclareMathAlphabet{\mathpzc}{OT1}{pzc}{m}{it}

\biboptions{sort&compress}

\journal{Physics Letters B}

\newcommand{\beq}{\begin{equation}}
\newcommand{\eeq}{\end{equation}}
\newcommand{\ba}{\begin{array}}
\newcommand{\ea}{\end{array}}
\newcommand{\bea}{\begin{align}}
\newcommand{\eea}{\end{align}}
\newcommand{\bi}{\begin{itemize}}
\newcommand{\ei}{\end{itemize}}
\newcommand{\ben}{\begin{enumerate}}
\newcommand{\een}{\end{enumerate}}
\newcommand{\bc}{\begin{center}}
\newcommand{\ec}{\end{center}}
\newcommand{\bl}{\begin{flushleft}}
\newcommand{\el}{\end{flushleft}}
\newcommand{\br}{\begin{flushright}}
\newcommand{\er}{\end{flushright}}

\begin{document}
\begin{CJK*}{UTF8}{gbsn}

%% Xiaobin Wang,0000-0002-2786-5296, 王晓斌
%% Zanbin Xing, 0000-0001-8910-6941, 幸瓒彬
%%  Lei Chang, 0000-0002-4339-2943, 常雷

\begin{frontmatter}

%\title{Projections for pion and kaon GPDs}
\title{$\,$\\[-7ex]\hspace*{\fill}{\normalsize{\sf\emph{Preprint no}. NJU-INP 094/24, USTC-ICTS/PCFT-24-48}}\\[1ex]
%Proton distribution functions}
%Linking hadron electromagnetic and gravitational form factors algebraically}
%
Kaon Distribution Functions from Empirical Information}

%% https://orcid.org/0000-0002-9104-9680
\author[UHe]{Zhen-Ni Xu (徐珍妮)%
       $^{\href{https://orcid.org/0000-0002-9104-9680}{\textcolor[rgb]{0.00,1.00,0.00}{\sf ID}},}$}
%\email{khepani.raya@dci.uhu.es}

\author[ECT]{Daniele Binosi%
    $\,^{\href{https://orcid.org/0000-0003-1742-4689}{\textcolor[rgb]{0.00,1.00,0.00}{\sf ID}}}$}
    
\author[USTC,PHCFT]{Chen Chen (陈晨)%
       $^{\href{https://orcid.org/0000-0003-3619-0670}{\textcolor[rgb]{0.00,1.00,0.00}{\sf ID}}}$}
%\email[]{chenchen1031@ustc.edu.cn}

\author[UHe]{\\Kh\'epani Raya%
       $^{\href{https://orcid.org/0000-0001-8225-5821}{\textcolor[rgb]{0.00,1.00,0.00}{\sf ID}},}$}
%\email{khepani.raya@dci.uhu.es}

\author[NJU,INP]{Craig D. Roberts%
       $^{\href{https://orcid.org/0000-0002-2937-1361}{\textcolor[rgb]{0.00,1.00,0.00}{\sf ID}},}$}
%\author[NJU,INP]{Craig D. Roberts}
%\email[]{cdroberts@nju.edu.cn}
%\ead{cdroberts@nju.edu.cn}

\author[UHe,CEA]{Jos\'e Rodr\'iguez-Quintero%
       $^{\href{https://orcid.org/0000-0002-1651-5717}{\textcolor[rgb]{0.00,1.00,0.00}{\sf ID}},}$}

\address[UHe]{Department of Integrated Sciences and Center for Advanced Studies in Physics, Mathematics and Computation, \href{https://ror.org/03a1kt624}{University of Huelva}, E-21071 Huelva, Spain.}

\address[ECT]{European Centre for Theoretical Studies in Nuclear Physics
            and Related Areas  (\href{https://ror.org/01gzye136}{ECT*}), Villa Tambosi, Strada delle Tabarelle 286, I-38123 Villazzano (TN), Italy}

\address[USTC]{Interdisciplinary Center for Theoretical Study, University of Science and Technology of China (\href{https://ror.org/04c4dkn09}{USTC}), Hefei, Anhui 230026, China}

\address[PHCFT]{Peng Huanwu Center for Fundamental Theory (\href{https://ror.org/01s9dmg43}{PCFT}), Hefei, Anhui 230026, China}

\address[NJU]{
School of Physics, \href{https://ror.org/01rxvg760}{Nanjing University}, Nanjing, Jiangsu 210093, China}
\address[INP]{
Institute for Nonperturbative Physics, \href{https://ror.org/01rxvg760}{Nanjing University}, Nanjing, Jiangsu 210093, China}

\address[CEA]{Irfu, CEA, Université de Paris-Saclay, 91191, Gif-sur-Yvette, France
\\[1ex]
%
%\hspace*{-8em}Email addresses:
{\rm
\href{mailto:chenchen1031@ustc.edu.cn}{chenchen1031@ustc.edu.cn} (CC);
\href{mailto:cdroberts@nju.edu.cn}{cdroberts@nju.edu.cn} (CDR);
\href{mailto:jose.rodriguez@dfaie.uhu.es}{jose.rodriguez@dfaie.uhu.es} (JRQ)
\\[1ex]
Date: 2024 November 21\\[-6ex]
%Date: 2024 October 29\\[-6ex]
%Date: 2024 October 22\\[-6ex]
}}
%%

%============================================================
\begin{abstract}
Using available information from Drell-Yan data on pion and kaon structure functions, an approach is described which enables the development of pointwise profiles for all pion and kaon parton distribution functions (DFs) without reference to theories of hadron structure.  The key steps are construction of structure-function-constrained probability-weighted ensembles of valence DF replicas and use of an evolution scheme for parton DFs that is all-orders exact.  The DFs obtained express qualitatively sound features of light-meson structure, \emph{e.g}., the effects of Higgs boson couplings into QCD and the size of heavy-quark momentum fractions in light hadrons.  In order to improve the results, additional and more precise data on the $u$-quark-in-kaon, ${\mathpzc u}^K$, to $u$-quark-in-pion, ${\mathpzc u}^\pi$, DF ratio would be necessary.  Of greater value would be extraction of ${\mathpzc u}^K$ alone, thereby avoiding inference from the ratio: currently, the data-based form of ${\mathpzc u}^K$ is materially influenced by results for ${\mathpzc u}^\pi$.
\end{abstract}

\begin{keyword}
continuum Schwinger function methods \sep
%Dyson-Schwinger equations \sep
Drell-Yan reactions \sep
emergence of mass \sep
meson structure functions \sep
Nambu-Goldstone bosons \sep
parton distribution functions
\end{keyword}

\end{frontmatter}
\end{CJK*}

%============================================================
\section{Introduction}
\label{s1}
The kaon was identified via a series of observations that began more than seventy years ago \cite{Rochester:1947mi, Brown:1949mj, Gell-Mann:1955ipe, Bardon:1958guy}.  Today, it is known to be a dichotomous system, \emph{viz}.\ both a bound state seeded by a light quark/antiquark and a strange antiquark/quark and, like the pion, first seen around the same time \cite{Lattes:1947mw}, a Nambu-Goldstone boson.  For instance, the valence content of the $K^+$ is $u \bar s$.  Moreover, the fascinating dichotomy can be resolved by exploiting the impact of interference between Nature's two known sources of mass generation; namely, Higgs boson couplings into QCD and emergent hadron mass (EHM) \cite{Roberts:2021nhw, Binosi:2022djx, Ding:2022ows, Ferreira:2023fva, Carman:2023zke, Raya:2024ejx}: pions and kaons are keys in the effort to elucidate EHM phenomena.

Data relating to pion structure are limited \cite{Horn:2016rip}, but the kaon case is dire; so, validating any picture of the kaon is especially difficult.  In fact, even after seventy years, practically nothing is known about kaon structure, \emph{e.g}., the precision of charged kaon elastic electromagnetic form factor, $F_K$, measurements is such that no objective result for the kaon charge radius is available \cite{Cui:2021aee}, the evolution of $F_K$ with squared momentum-transfer, $Q^2$, is uncertain \cite{Carmignotto:2018uqj}, and only eight points relating to in-kaon parton distribution functions (DFs) are available \cite{Badier:1980jq}.  These difficulties stem from the short lifetime of the kaon and its relatively low production rate.  The situation is expected to be improved in the coming decades by exploiting the capacity of high-luminosity, high-energy accelerators
\cite{Roberts:2021nhw, JlabTDIS2, Chen:2020ijn, Arrington:2021biu, Seitz:2023rqm, Metzger:2023vxl}.
%% pion lifetime 2.6033±0.0005  10^-8 s
%% kaon lifetime 1.2380±0.002 10^-8  s
%
In the interim, it is worth determining what can be inferred from existing kaon data and thereby highlight some of the insights that new data should bring.
Such information should also provide useful benchmarks for contemporary phenomenology and theory of Nambu-Goldstone boson structure, which, following recognition of its role as a clear window onto EHM, is receiving much attention -- see, \emph{e.g}., Refs.\,\cite{Kock:2020frx, Cui:2020tdf, Cui:2020dlm, Lin:2020ssv, Xie:2021ypc, Chavez:2021koz, dePaula:2022pcb, Pasquini:2023aaf, Wang:2023bmk, Good:2023ecp, Chang:2024rbs, Alexandrou:2024zvn}.

\section{DF Evolution and the Hadron Scale}
\label{s2}
To be concrete, we will speak of the $K^+$.  Apart from changing $u\to \bar u$, $\bar s \to s$, all remarks pertain equally to $K^-$.  So, denote the valence DFs of the $K^+$ by ${\mathpzc q}^K(x;\zeta)$, where ${\mathpzc q} = {\mathpzc u}, \bar{\mathpzc s}$, $x$ is the light-front fraction of the kaon's four-momentum carried by the indicated valence species, and $\zeta$ is the scale at which this fraction is resolved.  Any such DF is fully characterised by its Mellin moments ($m\in \mathbb Z^>$):
\begin{equation}
\langle x^m \rangle_{{\mathpzc q}^K}^\zeta
= \int_0^1 dx\, x^m {\mathpzc q}^K(x;\zeta)\,,
\end{equation}
with $\langle x^0 \rangle_{{\mathpzc q}^K}^\zeta = 1$ owing to baryon number conservation.

The relation between DFs at different resolving scales is given by the DGLAP equations \cite{Dokshitzer:1977sg, Gribov:1971zn, Lipatov:1974qm, Altarelli:1977zs}.  Herein, we exploit the all-orders (AO) scheme detailed in Ref.\,\cite{Yin:2023dbw}, which is built upon the following two propositions.
(\emph{I}) There is an effective charge, $\alpha_{1\ell}(k^2)$, of the type explained in Refs.\,\cite{Grunberg:1980ja, Grunberg:1982fw} and reviewed in Ref.\,\cite{Deur:2023dzc}, that, when used to integrate the leading-order perturbative DGLAP equations, defines an evolution scheme for \emph{every} parton DF that is all-orders exact.
The pointwise form of $\alpha_{1\ell}(k^2)$ is largely immaterial.  Nonetheless, the process-independent QCD running coupling calculated in Refs.\ \cite{Cui:2019dwv} has all required properties.
(\emph{II}) There is a scale, $\zeta_{\cal H}$, whereat all properties of a given hadron are carried by its valence degrees-of-freedom.  At $\zeta_{\cal H}$, DFs associated with glue and sea quarks are zero.
Working with the charge discussed in Refs.\,\cite{Cui:2019dwv, Deur:2023dzc, Brodsky:2024zev}, the value of the hadron scale is a prediction \cite{Cui:2021mom}:
$\zeta_{\cal H} = 0.331(2)\,{\rm GeV}$.
A consistent value, $\zeta_{\cal H} = 0.350(44)\,{\rm GeV}$, is obtained from an analysis of lattice-QCD (lQCD) results relating to the pion valence quark DF \cite{Lu:2023yna}.

The AO approach defines a nonperturbative extension of DGLAP evolution.  It has successfully been used in many applications, \emph{e.g}., 
delivering unified predictions for pion fragmentation functions \cite{Xing:2023pms} and all pion, kaon, and proton DFs \cite{Cui:2020tdf, Chang:2022jri, Lu:2022cjx, Cheng:2023kmt, Yu:2024qsd},
a viable species separation of nucleon gravitational form factors \cite{Yao:2024ixu}, 
and insights into quark and gluon angular momentum contributions to the proton spin \cite{Yu:2024ovn}.
%See also Refs.\,\cite{dePaula:2022pcb, Albino:2022gzs} for additional applications.

Within the AO scheme,
\begin{equation}
\langle x \rangle_{{\mathpzc u}^K}^{\zeta_{\cal H}} + \langle x \rangle_{\bar {\mathpzc s}^K}^{\zeta_{\cal H}} = 1\,;
\end{equation}
hence $\bar {\mathpzc s}^K(x;\zeta_{\cal H}) = {\mathpzc u}^K(1-x;\zeta_{\cal H}) $ and
\begin{equation}
{\mathpzc K}(x;\zeta_{\cal H}) = \tfrac{1}{2}
[{\mathpzc u}^K(x;\zeta_{\cal H})+\bar {\mathpzc s}^K(x;\zeta_{\cal H})]
= {\mathpzc K}(1-x;\zeta_{\cal H})\,.
\label{DFcalK}
\end{equation}
Working with this symmetric DF, one may capitalise on both the Mellin moment physical bounds and recursion relation presented in Ref.\,\cite{Cui:2022bxn}.

%Obtained using continuum Schwinger function methods (CSMs),
Predictions for hadron scale valence DFs in the pion and kaon are available \cite{Cui:2020tdf}.  For practical purposes, they can be interpolated using the following function:
\begin{equation}
\label{qDF1}
{\mathpzc q}^{M}(x;\zeta_{\cal H})
= n_0 \ln \left[ 1 + x^2 (1-x)^2 /\rho_M^2\right]
[ 1  \pm \gamma_M (1-2x)]\,,
\end{equation}
$M=\pi, K$, where $n_0=n_0(\rho_M)$ ensures $\langle x^0 \rangle_{{\mathpzc q}^M}^\zeta = 1$,
$\gamma_\pi=0$,
and, in the kaon, one has ``$+\gamma_K$'' for the $u$ quark and ``$-\gamma_K$'' for the $\bar s$.  Equally good interpolations are obtained using
\begin{equation}
\label{qDF2}
{\mathpzc u}^{M}(x;\zeta_{\cal H})
= \tilde n_0 \ln \left[ 1 + x^2 (1-x)^2
(1+\tilde\gamma_M^2 x^2 (1-x)^4)/\tilde \rho_M^2 \right]\,,
\end{equation}
with $\tilde n_0=n_0(\tilde\rho_M,\tilde\gamma_M)$,
$\tilde \gamma_\pi=0$,
$\bar {\mathpzc s}^{K}(x;\zeta_{\cal H}) = {\mathpzc u}^{K}(1-x;\zeta_{\cal H})$.
Both forms respect the $x\simeq 0,1$ endpoint behaviour predicted by continuum analyses of the meson bound state problem that preserve an identifiable connection with a gluon exchange interaction between valence degrees of freedom \cite[Sec.\,5.A]{Roberts:2021nhw}.
Moreover, for any $\rho_M >0$, they express the EHM-induced dilation that is a characteristic feature of $\pi$, $K$ DFs \cite{Roberts:2021nhw, Ding:2022ows, Alexandrou:2024zvn}; and with $\gamma_K\neq 0$ (equivalently, $\tilde \gamma_K\neq 0$), the skewing introduced by Higgs boson couplings into QCD is preserved \cite{Cui:2020dlm}.
For our purposes, Eq.\,\eqref{qDF1} is easier to work with, but it is better to use Eq.\,\eqref{qDF2} when considering fragmentation functions \cite{Xing:2023pms}.  This is not an issue because one may readily map $(\rho_M,\gamma_M) \to (\tilde\rho_M,\tilde \gamma_M)$; so, pass between the two forms without information loss.

{\allowdisplaybreaks
Using Eq.\,\eqref{qDF1}, one has
\begin{equation}
\langle x^m \rangle_{{\mathpzc q}^K}^{\zeta_{\cal H}}(\rho_K,\gamma_K)
=\langle x^m \rangle_{{\mathpzc K}}^{\zeta_{\cal H}}
\pm \gamma_{K}(
\langle x^m \rangle_{{\mathpzc K}}^{\zeta_{\cal H}}
-2 \langle x^{m+1} \rangle_{{\mathpzc K}}^{\zeta_{\cal H}})\,,
\end{equation}
where the subscript ${\mathpzc K}$ indicates a moment of the symmetrised DF in Eq.\,\eqref{DFcalK}.  For $m=1$,
with kaon DF skewing measured via $\delta_K = 1/2 - \langle x \rangle_{{\mathpzc u}_K}^{\zeta_{\cal H}} = \langle x \rangle_{\bar{\mathpzc s}_K}^{\zeta_{\cal H}}-1/2$,
\begin{equation}
\gamma_K = 2 \delta_K/[4 \langle x^2 \rangle_{{\mathpzc K}}^{\zeta_{\cal H}}-1]\,.
\end{equation}
}

\section{DFs at Empirical Scales}
An interpretation of experimental data in terms of parton DFs is only possible at resolving scales $\zeta\gg \zeta_{\cal H}$.  In this connection, the strength of the AO scheme becomes manifest because one finds \cite[Eq.\,(9)]{Yin:2023dbw}:
\begin{equation}
\langle x^m\rangle_{\mathpzc K}^\zeta/\langle x^m\rangle_{\mathpzc K}^{\zeta_{\cal H}}
 = [\langle 2 x \rangle_{\mathpzc K}^\zeta]^{\gamma_0^m/\gamma_0^1}\,,
 \label{EvolveAO}
\end{equation}
where
\begin{align}
\gamma_0^n
&= -\frac 4 3 \left[ 3 + \frac{2}{(n+1)(n+2)} - 4 \sum_{k=1}^{n+1} \frac 1 k  \right] \,, 
\label{eq:Dfn}
\end{align}
so $\gamma_0^0=0$, and all other anomalous dimensions may be found in Ref.\,\cite[Eq.\,(6)]{Yin:2023dbw}.  Equation~\eqref{EvolveAO} is the statement that given the symmetrised valence quark DF in Eq.\,\eqref{DFcalK}, then its pointwise form at any $\zeta > \zeta_{\cal H}$ is completely determined by its $m=1$ moment at that scale.  It is further worth recording that, ignoring quark current-mass dependence in the splitting kernels, which is a good approximation for $\pi$ and$K$ mesons \cite{Cui:2020tdf}:
\begin{equation}
\forall \zeta>\zeta_{\rm H} \; | \; \langle 2 x \rangle_{\mathpzc K}^\zeta = \langle 2 x \rangle_{{\mathpzc u}_\pi}^\zeta = \langle x \rangle_{{\mathpzc q}_K}^\zeta/\langle x \rangle_{{\mathpzc q}_K}^{\zeta_{\cal H}}
\; < 1 
\end{equation}
and
\begin{equation}
 \langle x^m\rangle_{{\mathpzc q}^M}^\zeta /\langle x^m\rangle_{{\mathpzc q}^M}^{\zeta_{\cal H}}
 =  [\langle 2 x \rangle_{{\mathpzc u}^\pi}^\zeta]^{\gamma_0^m/\gamma_0^1}\,.
 \label{EvolveAOG}
\end{equation}

The following function is a commonly used parametrisation of valence DFs at $\zeta > m_p$, where $m_p$ is the proton mass:
\begin{equation}
{\mathpzc q}^M(x;\zeta) = N_0(\alpha,\beta,c^M_q) \, x^{\alpha-1} (1-x)^\beta \left( 1 + c^M_q \, x^2 \right)
\,. \label{qMparam}
\end{equation}
Here, $N_0$ ensures unit normalisation for the $m=0$ DF moment and the parameters $(\alpha,\beta,c_q^M)$ depend on $\zeta$.  It is straightforward to establish that
\begin{align}
c^M_{q,\zeta} = -\frac{(2+\alpha+\beta) (3 + \alpha + \beta) (\alpha - (1+\alpha + \beta) \langle x \rangle_{{\mathpzc q}^M}^\zeta) }
{\alpha (1+\alpha) (2+\alpha- (3+\alpha+\beta) \langle x \rangle_{{\mathpzc q}^M}^\zeta)}\,.
\label{cMval}
\end{align}

Under flavour-independent nonsinglet AO evolution, $\alpha$ and $\beta$ evolve separately, but in the same way for all mesons and from the same starting point, \emph{i.e}., the values $\alpha(\zeta)$, $\beta(\zeta)$ are each generated from hadron scale DFs with $\alpha(\zeta_{\cal H}) = 3$, $\beta(\zeta_{\cal H})=2$ -- see Eq.\,\eqref{qDF1}.
Hence, ${\mathpzc u}^K/{\mathpzc u}^\pi$ is a nonzero finite number at $x=0,1$:
\begin{subequations}
\label{endpoints}
\begin{align}
\left.{\mathpzc u}^K/{\mathpzc u}^\pi\right|_{x=0}^\zeta & =
\frac{2 + \alpha - (3+\alpha+\beta) \langle x \rangle_{{\mathpzc u}^K}^\zeta}
{2 + \alpha - (3+\alpha+\beta) \langle x \rangle_{{\mathpzc u}^\pi}^\zeta}\,,\\
\left.{\mathpzc u}^K/{\mathpzc u}^\pi\right|_{x=1}^\zeta & =
\frac{\alpha (4 + 2\alpha +\beta) - (3+\alpha+\beta)(2+2\alpha+\beta)\langle x \rangle_{{\mathpzc u}^K}^\zeta}
{\alpha (4 + 2\alpha +\beta) - (3+\alpha+\beta)(2+2\alpha+\beta) \langle x \rangle_{{\mathpzc u}^\pi}^\zeta}\,.
\end{align}
\end{subequations}
It should be stressed that Eqs.\,\eqref{endpoints} are only applicable on some domain $\zeta > m_p$; namely, when Eq.\,\eqref{qMparam} provides a good representation of the pointwise form of the valence quark DFs.

Given that the light-front momentum fraction carried by valence quarks vanishes as $m_p/\zeta\to 0$, then
$\left.{\mathpzc u}_K/{\mathpzc u}_\pi\right|_{x=0}\to 1$ on this scale domain: singlet DFs become dominant on $x\simeq 0$.
Furthermore, $\left.{\mathpzc u}^K/{\mathpzc u}^\pi\right|_{x=1}<1$ for $\langle x \rangle_{{\mathpzc u}^K}^\zeta< \langle x \rangle_{{\mathpzc u}^\pi}^\zeta$, which is predicted by all studies with a sound implementation of Higgs boson couplings into QCD \cite{Cui:2020dlm}.
Both these predictions are compatible with the (sparse) available data \cite{Badier:1980jq}.
An illustration is readily extracted from the DFs calculated in Ref.\,\cite{Cui:2020tdf}, the $\zeta=\zeta_5=5.2\,$GeV results for which may be expressed using Eq.\,\eqref{qMparam} with $\alpha = 0.88$, $\beta = 2.52$ and
$\langle x \rangle_{{\mathpzc u}^K}^{\zeta_{\cal H}} = 0.47$,
$\langle 2 x \rangle_{{\mathpzc u}^\pi}^{\zeta_{5}} = 0.44$, to find
\begin{equation}
\left.{\mathpzc u}^K/{\mathpzc u}^\pi\right|_{x=0}^{\zeta_5}  =1.06\,,
\quad
\left.{\mathpzc u}^K/{\mathpzc u}^\pi\right|_{x=1}^{\zeta_5}  = 0.66\,.
\label{CuiRKrpi}
\end{equation}

\begin{figure}[t]
\vspace*{1ex}

\leftline{\hspace*{0.5em}{{\textsf{A}}}}
\vspace*{-4ex}
\centerline{\includegraphics[width=0.9\columnwidth]{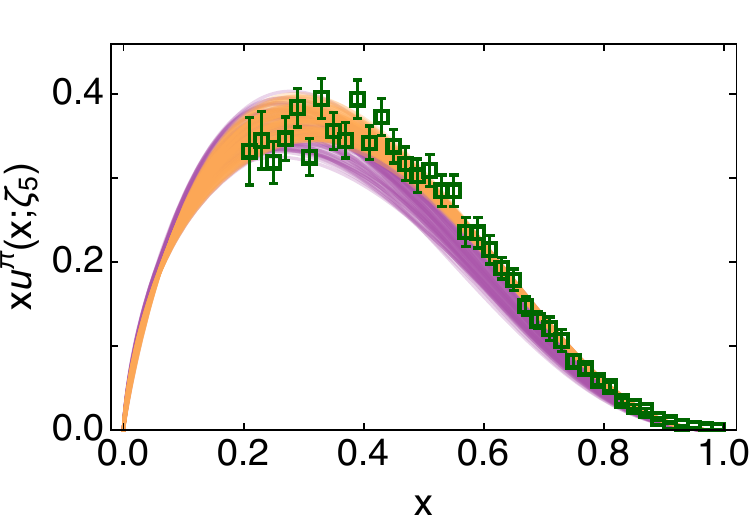}}
\vspace*{1ex}

\leftline{\hspace*{0.5em}{{\textsf{B}}}}
\vspace*{-1ex}
\centerline{\includegraphics[width=0.9\columnwidth]{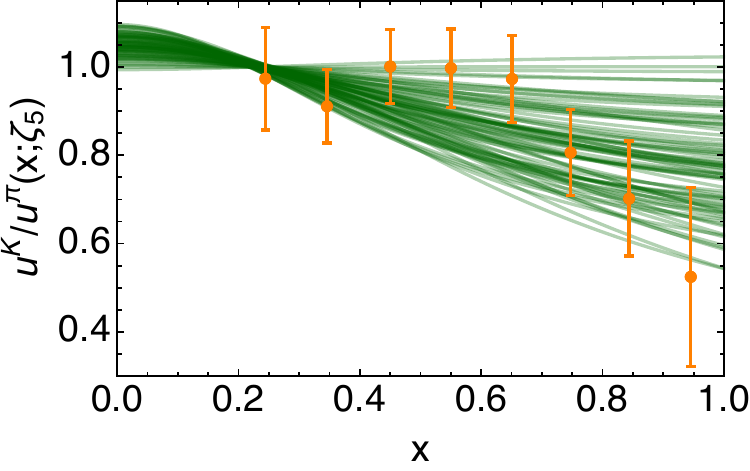}}

%\begin{tabular}{c}
%\includegraphics[width=0.85\columnwidth]{F2.pdf}
%\end{tabular}
%\vspace*{-0.25cm}
\caption{\label{F1}
Pion and kaon structure functions.
{\sf Panel A}.
Pion (purple) and kaon (orange) ensembles of valence $u$ quark DF replicas.
Points (green): analysis of Ref.\,\cite[E615]{Conway:1989fs} Drell-Yan data discussed in Refs.\,\cite{Aicher:2010cb, Chang:2014lva, Cui:2021mom}.
{\sf Panel B}.
Replicas for $K^- / \pi^-$ $u$ quark DF ratio (green curves).
Points (orange): analysis of Drell-Yan data in Ref.\,\cite{Badier:1980jq}.
}
\end{figure}

\section{Exploiting Available Data}
So far as pion and kaon DFs are concerned, two practicable data sets are available, \emph{viz}.\ $\pi$-induced Drell-Yan reaction cross-sections that may be understood with regard to ${\mathpzc u}^\pi(x;\zeta_5)$ \cite[E615]{Conway:1989fs} and Drell-Yan measurements that can be interpreted in terms of the $K^- / \pi^-$ structure function ratio at this same scale \cite{Badier:1980jq}, \emph{i.e}., ${\mathpzc u}^K(x;\zeta_5)/{\mathpzc u}^\pi(x;\zeta_5)$.
The analysis of E615 data discussed in Refs.\,\cite{Aicher:2010cb, Chang:2014lva} is displayed in Fig.\,\ref{F1}A -- the issues with other analyses are canvassed in Ref.\,\cite{, Cui:2021mom}; and the $K^- / \pi^-$ analysis from Ref.\,\cite{Badier:1980jq} is reproduced in Fig.\,\ref{F1}B -- these eight points represent all that is today known about the kaon structure function.
We now use this information to produce a data-driven estimate of pion and kaon DFs.

To begin, working with the E615 points drawn in Fig.\,\ref{F1}A, we use least-squares minimisation to determine the parameters that produce a best fit in the form of Eq.\,\eqref{qMparam}, with the result:
\begin{align}
[w^\pi_0]=\{\alpha_0,\beta_0,c_0^\pi\}
= \{ 0.792 , 2.71 , 2.49 \}\,.
\end{align}
%% {alpha -> 0.791665, beta -> 2.7143, cpi -> 2.49223}
It should be borne in mind that $\zeta=\zeta_5$ and $q=u = \bar d$.
Furthermore, the large-$x$ exponent, $\beta(\zeta_5) =2.71>2$, is consistent with QCD-based expectations \cite{Brodsky:1994kg, Yuan:2003fs, Holt:2010vj, Cui:2021mom, Cui:2022bxn, Lu:2022cjx}.

We now take the following steps.

(\emph{I}) Generate a set of new vectors $\{[w^\pi_j] \,| \, j=1, \ldots, N_w\}$, each element of which is distributed randomly around the best-fit result and accepted into a reference ensemble according to the probability
\begin{equation}
\label{EqProb}
{\mathpzc P}  = \frac{P(\chi^2;d)}{P(\chi_0^2;d)} \,, \;
P(y;d) = \frac{(1/2)^{d/2}}{\Gamma(d/2)} y^{d/2-1} {\rm e}^{-y/2}\,,
\end{equation}
where
\begin{equation}
\chi^2([w^\pi_j]) =
\frac{1}{\sigma}\sum_{i=1}^{M_{\rm E615}}\,
\frac{ [{\mathpzc u}^\pi(x_i;[w^\pi_j];\zeta_5)-u_i^{\rm E615}]^2}
{\delta_i^2};
\label{X2distribution}
\end{equation}
$\{(x_i,u_i \pm \delta_i) \, | \,  i =1, \ldots, M_{\rm E615} \}$ are the points associated with the E615 data;
and $\sigma$ is a rescaling factor that ensures $\chi^2([w^\pi_0]) = d-2$, making this the maximum of the probability density.

(\emph{II})
Recalling that $\pi$, $K$ valence quark DFs are characterised by the same values of $\alpha$, $\beta$, define
\begin{align}
R^{K/\pi}\left(x_j;[v_j^{K/\pi}];\zeta_5\right) & =
\frac{u^K(x_j;[w^j_K];\zeta_5)}{u^\pi(x_j;[w^j_\pi];\zeta_5)} \,,
\label{eq:RKoverPi}
\end{align}
which is the ratio depicted in Fig.\,\ref{F1}B.
Here, $[v_j^{K/\pi}] = \{\alpha_j,\beta_j, c_j^\pi, c_j^K\}$ is drawn from
$[w^\pi_j]=\{\alpha_j,\beta_j,c_j^\pi\}$,
$[w^K_j]=\{\alpha_j,\beta_j,c_j^K\}$.
The single unfixed parameter is $c_j^K$ in ${\mathpzc u}^K(x;\zeta_5)$.
Using Eq.\,\eqref{cMval}, it is determined by $\langle x \rangle_{{\mathpzc u}^{K}}^{\zeta_5}$.

(\emph{III})
For a given pion configuration, $[w^\pi_j]$, consider an associated kaon DF defined by $[w^K_j]$ with $c_j^K$ distributed randomly around a best fit value determined by the $K/\pi$ points drawn in Fig.\,\ref{F1}B and accepted into the reference ensemble with probability
\begin{equation}
{\mathpzc P}_{R^{K/\pi}} =
{\mathpzc P}_{R^{K/\pi}|{\mathpzc u}^\pi} {\mathpzc P}_{{\mathpzc u}^\pi} \,,
\end{equation}
where ${\mathpzc P}_{{\mathpzc u}^\pi}$ is the Step (\emph{I}) acceptance probability
and ${\mathpzc P}_{R^{K/\pi}|{\mathpzc u}^\pi}$ is calculated using Eq.\,\eqref{EqProb} with Eq.\,\eqref{X2distribution} amended so that  $R^{K/\pi}\left(x_j;[v_j^{K/\pi}];\zeta_5\right) $ becomes the first term within the numerator parentheses and the reference values and uncertainties are the $M_{K/\pi}=8$ points from Ref.\,\cite{Badier:1980jq}.
Unsurprisingly,
${\mathpzc P}_{R^{K/\pi}|{\mathpzc u}^\pi} \equiv {\mathpzc P}_{{\mathpzc u}^{K}|{\mathpzc u}^\pi}$,
so ${\mathpzc P}_{{\mathpzc u}^{K}} = {\mathpzc P}_{R^{K/\pi}}$; to wit,
given ${\mathpzc u}^\pi$, then the acceptance probability of a $K/\pi$ ratio is equivalent to that for ${\mathpzc u}^K$.

Beginning with $N_w=200$ and following Steps (I - III), we arrive at a $111$ member reference ensemble for ${\mathpzc u}^\pi(x;\zeta_5)$ and a $67$ member ensemble for $u^K(x;\zeta_5)$.  All members of both ensembles are displayed in Fig.\,\ref{F1}A; and the associated ensemble of $K/\pi$ ratios is drawn in Fig.\,\ref{F1}B.

A key reference result is the replica averaged value of
$\langle x \rangle_{{\mathpzc u}^\pi}^{\zeta_{5}} = 0.207(4)$.
Within mutual uncertainties, this value matches the CSM prediction \cite{Cui:2020tdf}: $0.20(2)$; and an average of mutually consistent lQCD results for this first moment \cite{Joo:2019bzr, Sufian:2019bol, Alexandrou:2021mmi, Gao:2022iex, Lu:2023yna}: $0.214(25)$.
%%uncertainty weighted average
%$\langle x \rangle_{{\mathpzc u}^\pi}^{\zeta_{5}} = 0.219(2)$.
%% Around[0.214, 0.025]
Of similar importance are the endpoint values of the ratio replicas drawn in Fig.\,\ref{F1}B:
\begin{equation}
\begin{array}{lcc}
& x\to 0  & x\to 1 \\
\displaystyle \frac{u^K(x;\zeta_5)}{u^\pi(x;\zeta_5)} & 1.050(24) & 0.77(11)
\end{array}\,.
\label{Rvalue}
\end{equation}
These results match the predictions in Ref.\,\cite{Cui:2020tdf}, reproduced in Eq.\,\eqref{CuiRKrpi}.
Plainly, more precise data are needed in order to reduce the $x\simeq 1$ uncertainty.

\begin{figure}[t]
\vspace*{-1ex}

    \centerline{\includegraphics[width=0.9\columnwidth]{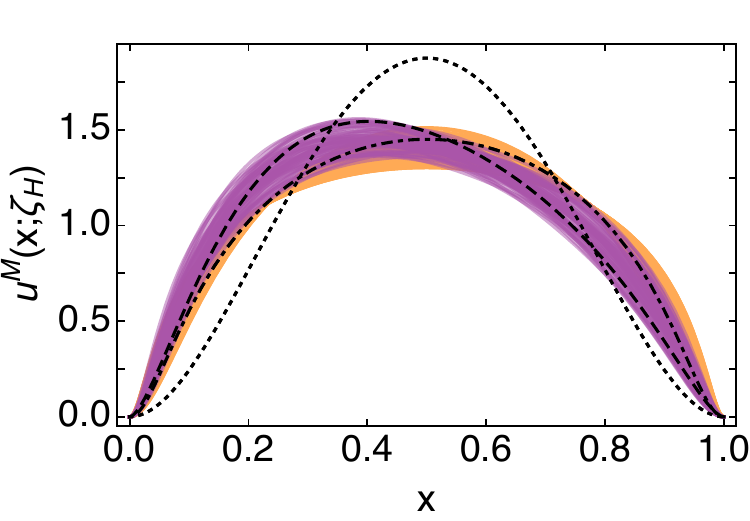}}
    \caption{\label{fig:plotszH}
    Hadron-scale evolved ensembles of pion (orange) and kaon (purple) valence $u$ quark DFs obtained following the procedure discussed in connection with Eq.\,\eqref{z5tozH}.
    Dot-dashed ($\pi$) and dashed $K$) curves are the predictions for these DFs in Ref.\,\cite{Cui:2020tdf}.
    Dotted curve is the scale-free DF ${\mathpzc q}_{\rm sf} (x) = 30 x^2(1-x)^2$.
    }
\end{figure}

Using the AO scheme, each of the ${\mathpzc u}^M(x;\zeta_5)$ DF replicas can be evolved back to $\zeta_{\cal H}$ using the replica value of $\langle x \rangle_{{\mathpzc u}^\pi}^{\zeta_{5}}$.
(We record the $\langle x \rangle_{{\mathpzc u}^\pi}^{\zeta_{5}}$ value associated with each accepted $R^{K/\pi}$.)
To be clear, beginning with Eq.\,\eqref{EvolveAOG}, one calculates the $\zeta_5\to \zeta_{\cal H}$ Mellin moments for a given replica, $j$, then minimises the following functional over $(\rho_M^j,\gamma_M^j)$:
\begin{equation}
\sum_{m=1}^{n_m} \left[
\langle x^m \rangle^{\zeta_{\cal H}}_{{\mathpzc u}^M(\rho_M^j,\gamma_M^j)}/
{\mathpzc M}_M^j(m)
 - 1 \right]^2,
 \label{z5tozH}
\end{equation}
where ${\mathpzc M}_M^j(m)$ is the $\zeta_5\to\zeta_{\cal H}$-evolved m$^{\rm th}$ Mellin moment of DF replica $j$.
($n_m=10$ is sufficient for accurate reconstruction and $\gamma_\pi^j\equiv 0$ in Eqs.\,\eqref{qDF1}, \eqref{qDF2}.)  The resulting ensembles of $\zeta_{\cal H}$ DFs are drawn in Fig.\,\ref{fig:plotszH}.
Using Eq.\,\eqref{qDF1}, the parameter averages and dispersions are
\begin{equation}
\rho_\pi = 0.051(15)\,, \quad
\rho_K = 0.054(13)\,, \gamma_K = 0.23(12)\,.
\label{zHvalues}
\end{equation}
If one instead chooses Eq.\,\eqref{qDF2}, then, naturally,
$\tilde\rho_\pi = \rho_\pi$, and
$\tilde\rho_K = 0.044(13)$, $\tilde\gamma_K=14(9)$.

A DF analogue of the asymptotic meson distribution amplitude, \emph{viz}.\ $\varphi_{\rm as}(x) = 6x(1-x)$ \cite{Lepage:1979zb, Efremov:1979qk, Lepage:1980fj},  is the function ${\mathpzc q}_{\rm sf} (x) = 30 x^2(1-x)^2$.
Compared with this profile, EHM is expressed in the dilation of each replica in the DF ensembles drawn in Fig.\,\ref{fig:plotszH}.
One notes from Fig.\,\ref{fig:plotszH} that the CSM predictions for pion and kaon valence DFs \cite{Cui:2020tdf} are consistent with the data-based inferences obtained herein, although the replica spreads, which owe to the imprecision of available data, diminish the significance of this correspondence.

Continuing with the discussion of Fig.\,\ref{fig:plotszH}, the peak location of each pion replica is $x=1/2$.  In contrast, the ${\mathpzc u}^K$ replicas peak at $x=0.42(4)$.  This relative shift reflects the impact of Higgs boson couplings into QCD on the lighter quark sector \cite{Cui:2020dlm}.  Notably, $0.5/0.42(4) = 1.20(11)$ and the kaon-to-pion leptonic decay constant ratio is $f_K/f_\pi = 1.19$ \cite{ParticleDataGroup:2024cfk}.  The similarity between the values of these ratios highlights their common origin and the fact that even though the $s$ quark current mass is $27$-times greater than the average light-quark mass \cite{ParticleDataGroup:2024cfk}, this is largely masked in kaon observables by the scale of EHM.

Regarding $\langle x \rangle_{{\mathpzc u}^K}^{\zeta_{\cal H}}$, one may (\emph{a}) calculate the hadron-scale valence $u$ quark momentum fraction by using AO evolution $\zeta_5 \to \zeta_{\cal H}$ on the first moment of each of the $67$ $\zeta_5$ replicas indicated by Fig.\,\ref{F1}.
Equally, one may (\emph{b}) follow the procedure discussed in connection with Eq.\,\eqref{z5tozH} to reconstruct ${\mathpzc u}^K(x;\zeta_{\cal H})$ and calculate the first moment of the result.
These two approaches yield mutually consistent values:
\begin{equation}
\begin{array}{lcc}
 & \mbox{(\emph{a})} & \mbox{(\emph{b})} \\
\langle x \rangle_{{\mathpzc u}^K}^{\zeta_{\cal H}} & 0.473(13) & 0.477(12)
\end{array}\,,
\end{equation}
which is a signal of the efficacy of our DF reconstruction procedure.

According to Ref.\,\cite{Holt:2010vj}, the $x=1$ value of the ratio ${\mathpzc u}^K(x;\zeta)/{\mathpzc u}^\pi(x;\zeta)$ is independent of $\zeta$.  At $\zeta_{\cal H}$, using Eq.\,\eqref{qDF1}:
\begin{equation}
R^{K/\pi}_1(\zeta_{\cal H})= \lim_{x\to 1} \frac{{\mathpzc u}^K(x;\zeta_{\cal H})}{{\mathpzc u}^\pi(x;\zeta_{\cal H})}
= \frac{n_0(\rho_K,\gamma_K)}{n_0(\rho_\pi,0)}  \frac{\rho_\pi^2}{\rho_K^2}  [1- \gamma_K]\,.
\label{ScaleInv}
\end{equation}
Inserting the DF reconstruction values from Eq.\,\eqref{zHvalues}, one obtains
$R^{K/\pi}_1(\zeta_{\cal H})=0.72(29)$, which matches the $\zeta_5$ value in Eq.\,\eqref{Rvalue}.
As another test of $R_1^{K/\pi}$ scale invariance,
we separately considered each calculated hadron-scale replica ${\mathpzc u}^{K,\pi}(x;\zeta_{\cal H})$,
evolved it to $\zeta=\zeta_2=2\,$GeV and $\zeta=\zeta_5$,
then calculated and compared the $x\simeq 1$ ratio at these two scales.
Over the $67$ independent replicas that could be used, the mean difference $R_1^{K/\pi}(\zeta_2)$ \emph{cf}.\ $R_1^{K/\pi}(\zeta_5)$ is just $1.6$\%.
These levels of agreement, with discrepancies far smaller than statistical uncertainties, are strong evidence in support of scale invariance for the endpoint ratio in Eq.\,\eqref{ScaleInv}.

%"The smallness of these differences, much smaller than statistical uncertainties,  makes apparent that Eq.(26)'s rhs is scale invariant when the parameters are obtained from properly evolved DFs."

\begin{figure}[t]
\vspace*{1.7ex}

\leftline{\hspace*{0.5em}{{\textsf{A}}}}
\vspace*{-2ex}
\centerline{\includegraphics[width=0.9\columnwidth]{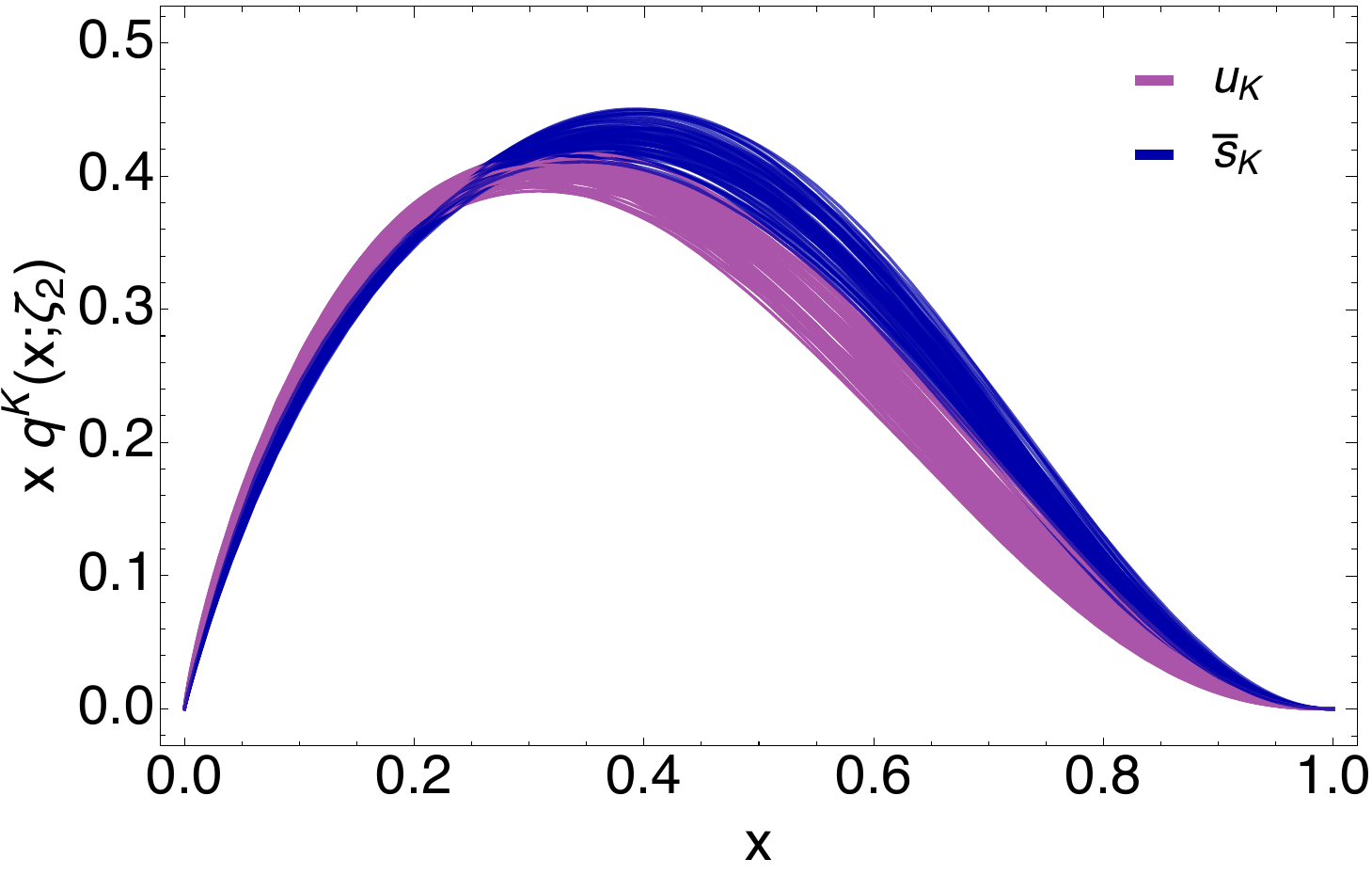}}
\vspace*{3ex}

\leftline{\hspace*{0.5em}{{\textsf{B}}}}
\vspace*{-2ex}
\centerline{\includegraphics[width=0.9\columnwidth]{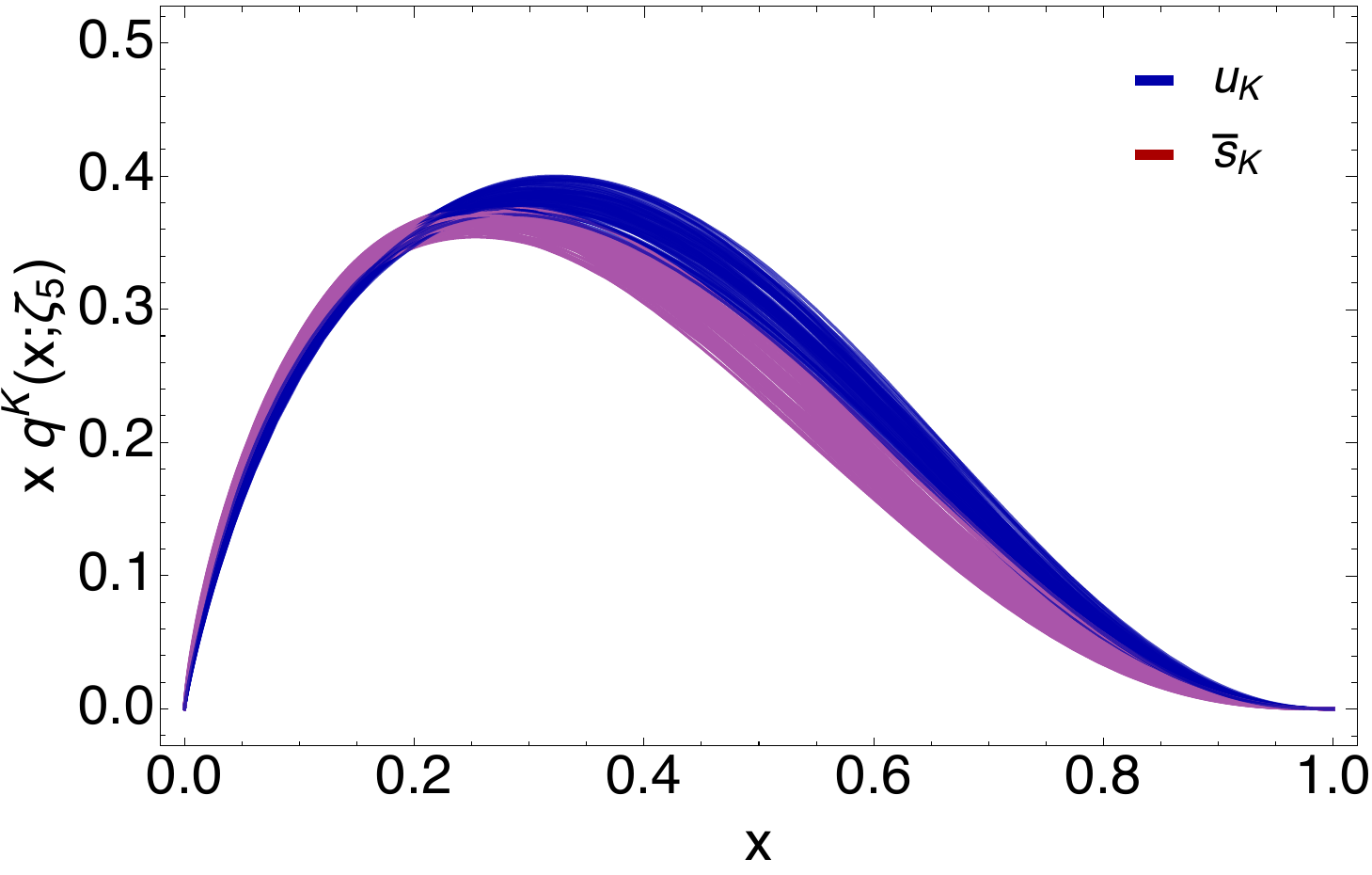}}

    \caption{\label{fig:usval}
    Data-driven ensemble predictions for valence quark DFs in the kaon:
    {\sf Panel A} -- $\zeta_2$; and {\sf Panel B} -- $\zeta_5$.}
\end{figure}

\section{Complete Set of Kaon and Pion DFs}
Having now built hadron scale valence DFs for the pion and kaon, \emph{viz}.\
${\mathpzc u}^{\pi, K}(x;\zeta_{\cal H})$ and $\bar {\mathpzc s}^{K}(x;\zeta_{\cal H})={\mathpzc u}^{K}(1-x;\zeta_{\cal H})$, all glue and sea DFs in these mesons can also be obtained via AO evolution.
In practice, we use the valence + singlet coupled system specified by Eqs.\,(5) in Ref.\,\cite{Yin:2023dbw}, with the hadron scale valence profiles as the initial surfaces, to produce $n_m$ Mellin moments of each DF at the desired scale $\zeta>\zeta_{\cal H}$.
As in Ref.\,\cite{Yin:2023dbw}, we employ quark-mass threshold functions to ensure that a given quark flavour only participates in evolution after the resolving scale exceeds
$\epsilon_q=\zeta_H+\delta_q$, with $\delta_u=\delta_d=0$ and (in GeV) $\delta_s=0.1$, $\delta_c=0.9$, $\delta_b=3.9$.
These threshold values are inferred from CSM predictions of dressed quark mass functions and their values at infrared momenta -- see, \emph{e.g}., Ref.\,\cite[Fig.\,2.5]{Roberts:2021nhw}.
With inputs based on a quark + interacting-diquark Faddeev equation picture \cite{Lu:2022cjx}, this approach to quark mass thresholds predicts a small but measurable $c+\bar c$ light-front momentum fraction ($\approx 1.3$\%) in the proton at $\zeta=\zeta_2$.
The pointwise form of each DF is subsequently obtained by following a procedure analogous to that discussed in connection with Eq.\,\eqref{z5tozH}, but using numerator Mellin moments associated with Eq.\,\eqref{qMparam}.

\begin{table}[t]
\centering
\begin{tabular}{c|cccc}
    & $N_0$ & $\alpha$ & $\beta$ & $c^K$   \\
\hline
    $\zeta_2$  &  &  &  & \\
\cline{1-1}
   $u$ & 2.48(23) & 0.88(3)& 2.42(12) & 1.38(27) \\
   $\bar{s}$ & 2.36(14) & 0.90(3)& 2.39(11) & 2.56(43) \\
\cline{2-5}
$u\bar{u},d\bar{d}$ & 0.0298(7) & -0.544(5) & 4.33(8) & 0 \\
$s\bar{s}$  & 0.0267(7) & -0.555(5) & 4.36(8) & 0 \\
$c\bar{c}$  & 0.0065(1) & -0.618(4) & 4.43(8) & 0 \\
$g$ & 0.384(11) & -0.566(5) & 3.31(8) & 0 \\
\hline
    $\zeta_5$ & & & & \\
\cline{1-1}
   $u$ & 2.22(18) & 0.79(3)& 2.67(12) & 1.20(24) \\
   $\bar{s}$ & 2.26(15) & 0.83(3)& 2.65(11) & 2.16(33) \\
\cline{2-5}
$u\bar{u},d\bar{d}$ & 0.0310(7) & -0.605(4) & 4.67(9) & 0 \\
$s\bar{s}$  & 0.0287(5) & -0.613(4) & 4.69(9) & 0 \\
$c\bar{c}$  & 0.0128(3) & -0.664(3) & 4.76(8) & 0 \\
$b\bar{b}$  & 0.0015(1) & -0.698(3) & 4.78(8) & 0 \\
$g$ & 0.313(7) & -0.642(4) & 3.60(8) & 0 \\
\hline
\end{tabular}
\caption{Used with Eq.\,\eqref{qMparam}, these coefficients deliver efficacious representations of all kaon DFs.
The uncertainties express the statistical variation over input valence ensembles.
For valence DFs, $N_0$ is fixed by baryon number conservation, but it is a free least-squares fitting parameter for sea and glue DFs.
\label{tab:params}}
\end{table}

\begin{table}[t]
\centering
\begin{tabular}{c|ccc|c}
    & $\langle x \rangle_{{\mathpzc q}^K}^{\zeta}$ & $\langle x \rangle_{{\cal S}_{\mathpzc q}^K}^{\zeta}$ & $\langle x \rangle_{{\mathpzc g}^K}^{\zeta}$
    & $\langle x \rangle_{{\mathpzc q}^\pi}^{\zeta}$
     \\ \hline
    $\zeta_2$ & & & \\ \cline{1-1}
    $u$ & 0.230(6)(10) & 0.028(2) &  & 0.241(5)(10) \\
    $d$ & 0 & 0.028(2) &   & 0.241(5)(10) \\
    $s$ & 0.252(6)(11) & 0.026(1) &   &   \\
    $c$ & 0 & 0.008(1) &   &   \\
    $b$ & 0 & 0 &  &   \\
    $g$ &   &   & 0.428(18) &   \\ \hline
    $\zeta_5$ & & & \\ \cline{1-1}
    $u$ & 0.197(5)(9) & 0.036(2) &  & 0.207(4)(9) \\
    $d$ & 0 & 0.036(2) &  & 0.207(4)(9) \\
    $s$ & 0.216(5)(9) & 0.034(2) &   &   \\
    $c$ & 0 & 0.019(1) &  &   \\
    $b$ & 0 & 0.003(1)  &  &   \\
    $g$ &  &   & 0.461(20) &   \\
\end{tabular}
\caption{Light-front momentum fractions for valence-, sea and glue (columns 1, 2, 3, respectively), at resolving scales $\zeta_2$ and $\zeta_5$, obtained from replicas in Fig.\,\ref{fig:plotszH} by AO evolution.
For valence degrees-of-freedom, the first uncertainty is statistical and the second expresses a $\pm 5$\% variation in $\zeta_{\cal H}$.
In all other cases, the statistical error is negligible.
For context, pion valence momentum fractions are given in the final column.  Ignoring flavour dependence in nonsinglet evolution equations, pion sea and glue fractions are the same as those in the kaon -- see Table~\ref{tab:momentslat}.
\label{tab:moments}}
\end{table}

\begin{table}[t]
\begin{center}
%\begin{tabular}{l|cc}
%    &  Here & Ref.\,\cite{Alexandrou:2024zvn} \\
%    \hline
%    $u+d$ & 0.286(12) & 0.260(9) \\
%    $s$ & 0.278(13) &  0.333(11) \\
%    $c$ & 0.008(1) &  0.024(17) \\
%    $u+d+s+c$ & 0.572(18) & 0.618(32)\\
%    $g$ & 0.428(18) &  0.408(61) \\ \hline
\begin{tabular}{c|cc|cc}
     &  empirical & & \cite[lQCD]{Alexandrou:2024zvn} &  \\
    \hline
    $M$ & $\pi$ & $K$ & $\pi$ & $K$ \\ \cline{1-5}
    $l$ & 0.538(15) & 0.286(12) & 0.499(55) & 0.317(19) \\
    $s$ & 0.026(01) & 0.278(13) & 0.036(15) & 0.339(11) \\
    $c$ & 0.008(01) & 0.008(01)  & 0.013(16) &  0.028(21) \\
    $q$ & 0.572(15) & 0.572(18) & 0.575(79) & 0.683(50) \\
    $g$ & 0.428(18) & 0.428(18) & 0.402(53) &  0.422(67) \\ \hline
\end{tabular}
\end{center}
\caption{Data-driven light-front momentum fractions calculated at $\zeta_2$ herein compared with available kindred results from lQCD \cite{Alexandrou:2024zvn}.
$l$ indicates the total fraction carried by light ($u$, $d$) quarks in the meson.
\label{tab:momentslat}}
%\caption{First Mellin moment for the gluon and singlet combinations of quark flavor kaon PDFs at the scale $\zeta_2$, expressing the momentum fraction carried by the separated flavors and glue, compared to the lattice results delivered in Ref.\,\cite{Alexandrou:2024zvn}.}
\end{table}

Employing this approach and working from the ensembles in Fig.\,\ref{fig:plotszH}, we obtain the $\zeta=\zeta_2, \zeta_5$ data-driven predictions for kaon valence DFs depicted in Fig.\,\ref{fig:usval}.  They are described by Eq.\,\eqref{qMparam} with the coefficients listed in Table~\ref{tab:params}.
These valence DFs yield the light-front momentum fractions listed in Table~\ref{tab:moments}, which express the fact that the $\bar{\mathpzc s}^K$ profile peaks at larger $x$ than ${\mathpzc u}^K$ owing to Higgs boson modulation of EHM.  Moreover, like nucleons, a measurable component of pion and kaon light-front momentum is carried by the $c+\bar c$ sea.

\begin{figure}[t]
\vspace*{0ex}

    \centerline{\includegraphics[width=0.95\columnwidth]{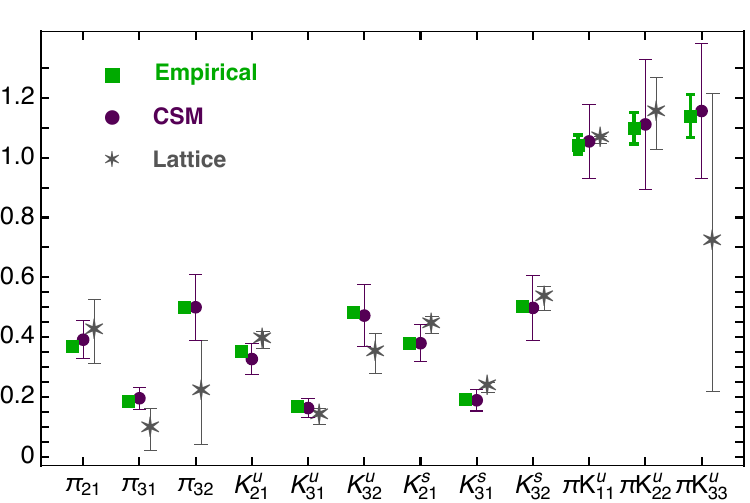}}
    \caption{\label{fig:plotsratio}
    Quark flavour-separated singlet Mellin moment ratios for the pion and kaon at $\zeta_2$.
    Legend.  $\pi_{ij} = \langle x^i \rangle_{{\mathpzc u}^\pi_S}/\langle x^j \rangle_{{\mathpzc u}_S^\pi}$,
    $K_{ij}^q = \langle x^i \rangle_{{\mathpzc q}_S^K}/\langle x^j \rangle_{{\mathpzc q}_S^K}$,
    $\pi K_{ij}^u = \langle x^i \rangle_{{\mathpzc u}_S^\pi}/\langle x^j \rangle_{{\mathpzc u}_S^K}$;
    ``Empirical'' -- results inferred herein from Refs.\,\cite{Conway:1989fs, Badier:1980jq}, ``CSM'' -- predictions from Ref.\,\cite{Cui:2020tdf}, ``Lattice'' -- lQCD results from Ref.\,\cite{Alexandrou:2020gxs, Alexandrou:2021mmi, AlexandrouBNL}.
    }
\end{figure}

The modest size of the relative shifts in the peak locations of pion, $u$-in-kaon and $\bar s$-in-kaon DFs is highlighted by Fig.\,\ref{fig:plotsratio} -- see the three rightmost points, which depicts a collection of Mellin moment ratios and compares the values inferred herein, from results reported in Refs.\,\cite{Conway:1989fs, Badier:1980jq}, with CSM \cite{Cui:2020tdf} and lQCD \cite{Alexandrou:2020gxs, Alexandrou:2021mmi, AlexandrouBNL} predictions for these same ratios.
A merit of such a comparison is that, \emph{e.g}., lQCD analyses can typically only deliver a few low-order moments, so objective conclusions can be drawn without reference to potentially subjective DF reconstruction using only a few moments.
Evidently, the CSM predictions \cite{Cui:2020tdf} are well aligned with the empirical results, whereas available lQCD values \cite{Alexandrou:2020gxs, Alexandrou:2021mmi, AlexandrouBNL} tend to lie low when higher $u$ quark moments are involved.  This may indicate an underestimate of the larger-$x$ support in the $u$ quark DFs computed using lQCD -- see also Table~\ref{tab:momentslat}.
It should be noted that the depicted lQCD results were obtained with a pion mass $m_\pi=0.26\,$GeV at a lattice spacing $a=0.093$fm.  No attempts were made to reach the physical pion mass or the continuum limit.
Related improvements are underway \cite{Alexandrou:2024zvn}. 

%% Constantia Alexandrou wrote on 20/11/2024 12:00 AM:
%% They are computed using one ensemble simulated at pion mass 260 MeV and lattice spacing 0.093 fm
%% Our new data for <x> used ensembles at the physical pion mass and are extrapolated to the continuum limit, see e-Print: 2405.08529 [hep-lat].   We find   <x>_\pi^u/<x>_kaon^u=0.87(16)
%%  I am attaching here for your information. As you can see  the continuum extrapolation  reduces the ratio but also increases its error. The blue point is at lattice spacing 0.08 fm. Having a fourth ensemble will help reduce the error on the final continuum result.
%%
% I believe it would be inconsistent to show only one point at the continuum limit in the plot.  However, I suggest  that it be  should be emphasised that the lattice results are not in the continuum limit.
%  Hopefully we will analyse the higher moments soon and then we can compare more meaningfully.

\begin{figure}[t]
\vspace*{1.7ex}

\leftline{\hspace*{0.5em}{{\textsf{A}}}}
\vspace*{-1ex}
\centerline{\includegraphics[width=0.9\columnwidth]{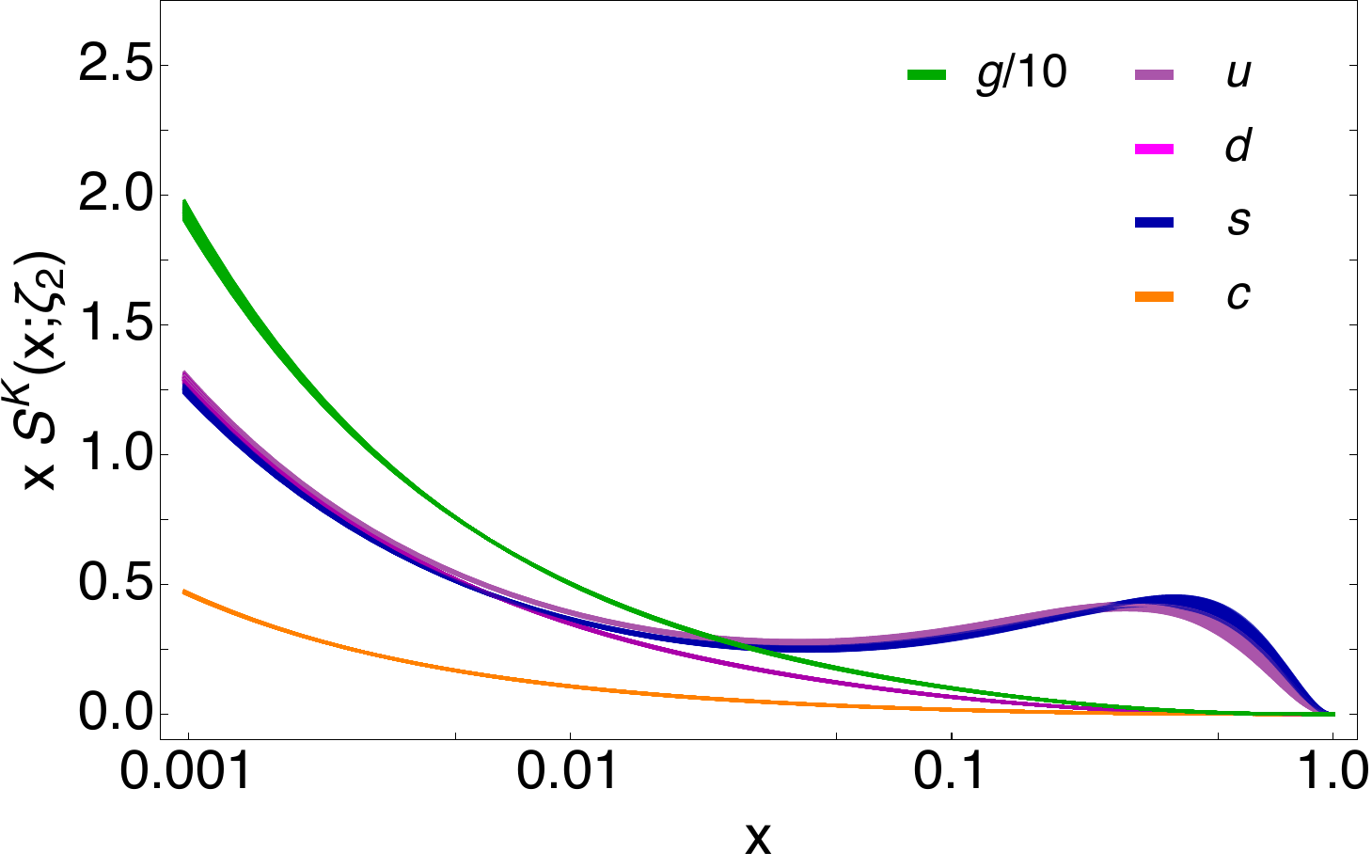}}
\vspace*{3ex}

\leftline{\hspace*{0.5em}{{\textsf{B}}}}
\vspace*{-1ex}
\centerline{\includegraphics[width=0.9\columnwidth]{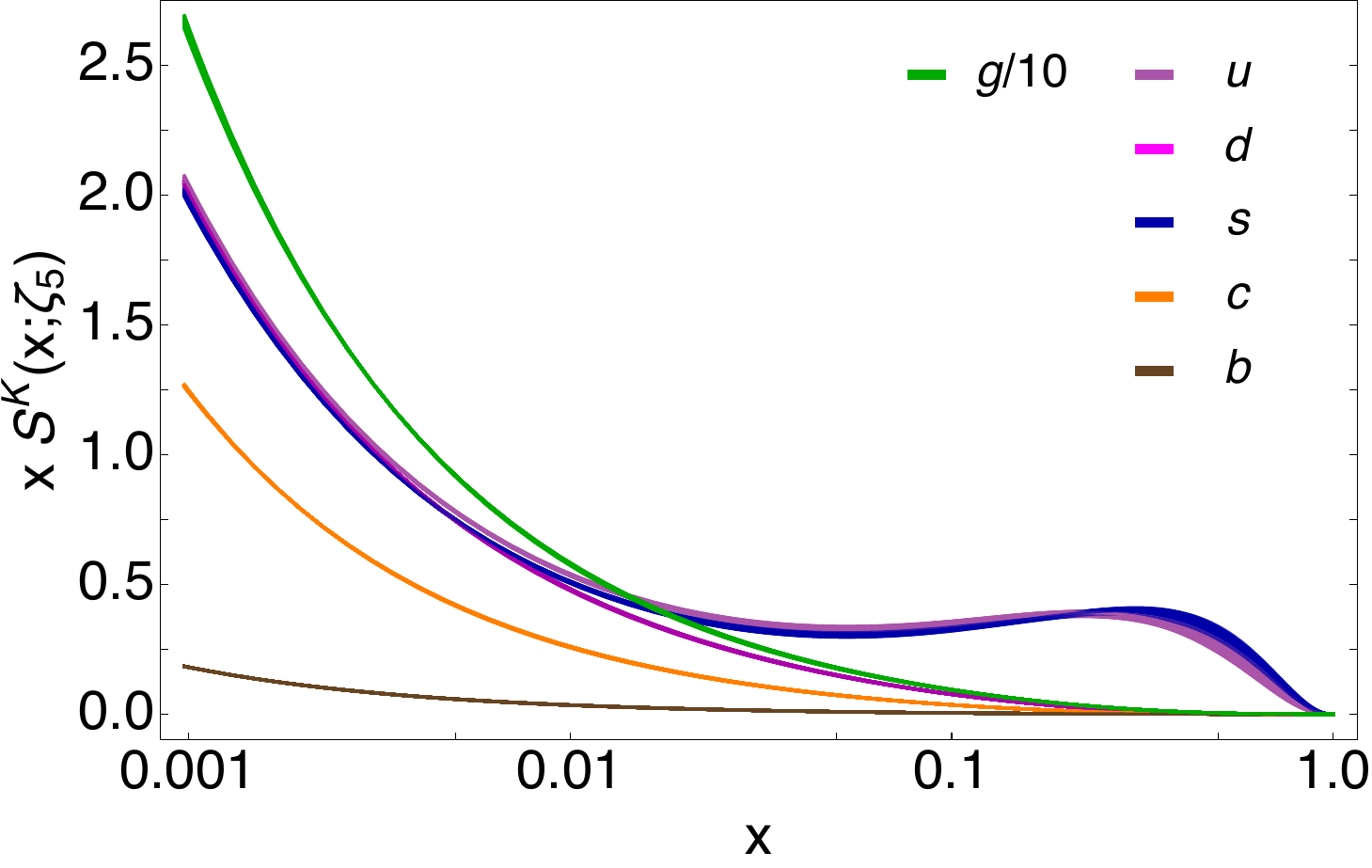}}

\caption{\label{fig:seaglue}
    Sea and glue DFs obtained via AO evolution of the kaon valence DF ensembles: {\sf Panel A} -- $\zeta_2$; and {\sf Panel B} -- $\zeta_5$.  $b+\bar b$ sea contributions only appear in Panel~B because $\zeta_2$ lies below the $b$ quark flavour threshold.}
\end{figure}

The kaon sea and glue profiles are drawn in Fig.\,\ref{fig:seaglue}.
Their pointwise forms are recovered by using Eq.\,\eqref{qMparam} and the appropriate coefficients from Table~\ref{tab:params}.
Notably, $\beta_{\rm sea} \approx \beta_{\rm glue} + 1 \approx \beta_{\rm valence}+2$, in accordance with QCD-based expectations \cite{Brodsky:1994kg}.
It is worth stressing that all DF ensembles drawn in Figs.\,\ref{fig:usval}, \ref{fig:seaglue} are determined solely by empirical information on ${\mathpzc u}^\pi$, ${\mathpzc u}^K$ and AO evolution.  No model or theory of meson structure has been assumed/employed.
The light-front momentum fraction carried by each species is listed in Table~\ref{tab:moments}.

We have already highlighted that the data-driven predictions for pion and kaon DFs delivered herein are consistent with modern CSM predictions \cite{Cui:2020tdf}.
It is also interesting to compare them with available information from lQCD studies.
In recent work, Ref.\,\cite{Alexandrou:2024zvn} used three gauge-field ensembles generated with $2+1+1$-flavours of Wilson fermions at physical quark current masses and subsequently extrapolated to the continuum limit to provide an array of species separated light-front momentum fractions in the pion and kaon.
They are reproduced in Table~\ref{tab:momentslat} and compared with like results from our analysis.
Within mutual uncertainties, which are sizeable for the lQCD results ($\approx 20$\% for pion and $\approx 10$\% for kaon, when the signal is strong), the level of agreement is fair and there are no striking discrepancies, especially when one accounts for the central-value violation of the momentum sum rule in the lQCD study.
%Nothing more can be said before uncertainties on both structure function data analyses and the lQCD results are significantly reduced.
Nothing more can be said before uncertainties on the lQCD results are reduced.

It is worth reiterating that the light-front momentum fractions listed in Table~\ref{tab:moments} do not depend on details concerning the $x$-dependence of ${\mathpzc u}^{\pi,K}(x,\zeta_{\cal H})$.  Instead, their values are completely determined by the fact that $\langle x \rangle_{{\mathpzc u}^\pi}^{\zeta_{\cal H}}+\langle x \rangle_{\bar{\mathpzc s}^\pi}^{\zeta_{\cal H}} \equiv 1$ and capitalisation on this feature via AO evolution.
Consequently, the values in Table~\ref{tab:moments}\,--\, columns 2 and 3 are predictions of the AO method, as are the total quark and glue momentum fractions reported in Table~\ref{tab:momentslat}.  
Under AO evolution, the momentum fractions carried by sea and glue are hadron independent \cite{Lu:2022cjx, Yin:2023dbw}, unless the nonsinglet splitting kernels are modified as discussed now.

In using AO evolution hitherto, we have neither implemented Pauli blocking \cite{Chang:2022jri, Lu:2022cjx} nor quark current-mass dependence in evolution of valence DFs \cite[Sec.\,7.3]{Cui:2020tdf}.  Consequently, the glue and flavour-summed quark momentum fractions listed in Tables~\ref{tab:moments}, \ref{tab:momentslat} are the same in the pion and kaon.  Were we to follow Ref.\,\cite[Sec.\,7.3]{Cui:2020tdf} in implementing mass dependence in valence DF evolution, then, \emph{e.g}., valence $\bar s$ quarks would produce less gluons than valence $u$ quarks and gluon splitting would produce less $s+\bar s$ pairs than light-quark pairs.  In consequence, we would obtain
\begin{align}
\label{ModGlue}
\langle x \rangle_{g^K}^\zeta =
\left\{
\begin{array}{rl}
0.4249 (2) (180) & \zeta=2\,\textrm{GeV} \\
0.4582 (2) (200) & \zeta=5\,\textrm{GeV}
\end{array}
\right. \;.
\end{align}
(Uncertainties are statistical and $\zeta_{\cal H} \to (1\pm 0.05)\zeta_{\cal H}$ systematic.)
Compared with the results in Table~\ref{tab:moments}, the differences are within existing uncertainties.  This explains why we have neglected the effect herein.  Similar statements hold for Pauli blocking in the $\pi$ and $K$ -- see the analysis in Ref.\,\cite{Yin:2023dbw}.

It is worth recording here that a suggestion \cite{Chen:2016sno} that the pion contains much more glue than the kaon was predicated on early fits to $\pi$-nucleon Drell-Yan data \cite{Gluck:1999xe}.  Relative to contemporary CSM predictions and the analysis described herein, those fits placed too little momentum with the pion's valence degrees of freedom, thereby misguiding the analysis in Ref.\,\cite{Chen:2016sno}.  The modern picture indicates \cite{Cui:2020tdf} that the in-kaon gluon momentum fraction is just $1$\% less than that in the pion -- compare Eq.\,\eqref{ModGlue} with Table~\ref{tab:momentslat}\,--\,Row~5.

\section{Summary and Perspective}
We described an approach to the reconstruction of pointwise profiles for all pion and kaon parton distribution functions (DFs) from valence parton DFs inferred via phenomenological analyses of relevant data.  
The key elements are data-constrained probability-weighted ensembles of valence DF replicas and the all-orders (AO) evolution scheme.  The latter is crucial because it enables valence DFs to be brought together at a common hadron scale and thereat used as the initial value profiles in coupled evolution equations for all parton species.

Herein, one aim was to reveal what could be learnt from existing data (eight imprecise points) on the $u$-quark-in-kaon to $u$-quark-in-pion structure function ratio, $R^{K/\pi}={\mathpzc u}^K(x)/{\mathpzc u}^\pi(x)$, when combined with available pion valence DF information derived from $\pi$-nucleon Drell-Yan measurements.  Regarding the valence DFs and considering $K^+$ as the exemplar, the ensembles obtained express qualitatively sound features of kaon structure, such as the size of the relative shift in $u$ and $\bar s$ DF peak locations caused by Higgs boson couplings into QCD [Fig.\,\ref{fig:usval}] and reasonably tight constraints on the associated light-front momentum fractions [Table~\ref{tab:moments}].  Similar statements are applicable to all singlet DFs [Figs.\,\ref{fig:plotsratio}, \ref{fig:seaglue}].  Of some interest are the results that the $c+\bar c$ sea carries approximately 2\% of the kaon's light-front momentum at resolving scale $\zeta_5=5.2\,$GeV, with $\approx 0.3$\% carried by $b+\bar b$ [Table~\ref{tab:moments}].

An open question is what data should be collected in order to aid in delivering more refined statements about kaon and pion structure.  Most simply, more, precise data on $R^{K/\pi}$ would be valuable.  At present,
the ${\mathpzc u}^K$ and $\bar{\mathpzc s}^K$ valence DF ensembles obtained therefrom overlap on large $x$ domains; so, the true size of EHM modulation by Higgs boson couplings into QCD is unclear.  More valuable still would be extraction of $u^K$ alone, \emph{i.e}., avoiding inference from $R^{K/\pi}$.  At present, the extracted form of ${\mathpzc u}^K$ is too much influenced by results for, or assumptions about, ${\mathpzc u}^\pi$.  Many calculations have shown that the same $R^{K/\pi}$ profile is produced by distinctly different, yet internally consistent, results for ${\mathpzc u}^{K,\pi}$  -- see, \emph{e.g}., Ref.\,\cite[Fig.\,12]{Cui:2020tdf}, which contrasts a CSM prediction of the ratio with a lQCD result \cite{Lin:2020ssv}.  Such data is anticipated in the foreseeable future \cite{Roberts:2021nhw, JlabTDIS2, Chen:2020ijn, Arrington:2021biu, Seitz:2023rqm, Metzger:2023vxl}.

%Approach demonstrated to be viable ... data + AO evolution yields sound reconstruction of pion and kaon DFs.

%When lQCD results become more precise, perhaps they could be included in the mix. ... don't think we want to say this.

\medskip

\noindent\textbf{Acknowledgments}.
%; and to D.~Binosi for supplying the CSM curve reproduced in the lower panel therein.
%
Work supported by:
National Natural Science Foundation of China (grant no.\ 12135007, 12247103);
Spanish Ministry of Science, Innovation and Universities (MICIU grant no.\ PID2022-140440NB-C22);
%
%Junta de Andaluc{\'{\i}}a (grant no.\ P18-FR-5057);
%
Program d'Alembert,
Project AST22$\_00001\_$X, funded by NextGenerationEU and ``Plan de Recuperaci\'on, Transformaci\'on y Resiliencia y Resiliencia'' (PRTR) from MICIU and Junta de Andaluc{\'{\i}}a;
and completed in part at Institut Pascal, Universit\'e Paris-Saclay, with the support of the program ``Investissements d'avenir'' ANR-11-IDEX-0003-01.

\medskip
\noindent\textbf{Data Availability Statement}. This manuscript has no associated data or the data will not be deposited. [Authors' comment: All information necessary to reproduce the results described herein is contained in the material presented above.]

\medskip
\noindent\textbf{Declaration of Competing Interest}.
The authors declare that they have no known competing financial interests or personal relationships that could have appeared to influence the work reported in this paper.

%\clearpage
%%\bibliographystyle{elsarticle-num-names}
%%\bibliography{../../../CollectedBiB}

\begin{thebibliography}{70}
\providecommand{\natexlab}[1]{#1}
\providecommand{\url}[1]{\texttt{#1}}
\providecommand{\urlprefix}{URL }
\expandafter\ifx\csname urlstyle\endcsname\relax
  \providecommand{\doi}[1]{doi:\discretionary{}{}{}#1}\else
  \providecommand{\doi}[1]{doi:\discretionary{}{}{}\begingroup
  \urlstyle{rm}\url{#1}\endgroup}\fi
\providecommand{\bibinfo}[2]{#2}

\bibitem[{Rochester and Butler(1947)}]{Rochester:1947mi}
\bibinfo{author}{G.~D. Rochester}, \bibinfo{author}{C.~C. Butler},
  \bibinfo{title}{{Evidence for the Existence of New Unstable Elementary
  Particles}}, \bibinfo{journal}{Nature} \bibinfo{volume}{160}
  (\bibinfo{year}{1947}) \bibinfo{pages}{855--857}.

\bibitem[{Brown et~al.(1949)Brown, Camerini, Fowler, Muirhead, Powell, and
  Ritson}]{Brown:1949mj}
\bibinfo{author}{R.~Brown}, \bibinfo{author}{U.~Camerini},
  \bibinfo{author}{P.~H. Fowler}, \bibinfo{author}{H.~Muirhead},
  \bibinfo{author}{C.~F. Powell}, \bibinfo{author}{D.~M. Ritson},
  \bibinfo{title}{{Observations With Electron Sensitive Plates Exposed to
  Cosmic Radiation}}, \bibinfo{journal}{Nature} \bibinfo{volume}{163}
  (\bibinfo{year}{1949}) \bibinfo{pages}{82}.

\bibitem[{Gell-Mann and Pais(1955)}]{Gell-Mann:1955ipe}
\bibinfo{author}{M.~Gell-Mann}, \bibinfo{author}{A.~Pais},
  \bibinfo{title}{{Behavior of neutral particles under charge conjugation}},
  \bibinfo{journal}{Phys. Rev.} \bibinfo{volume}{97} (\bibinfo{year}{1955})
  \bibinfo{pages}{1387--1389}.

\bibitem[{Bardon et~al.(1958)Bardon, Lande, Lederman, and
  Chinowsky}]{Bardon:1958guy}
\bibinfo{author}{M.~Bardon}, \bibinfo{author}{K.~Lande}, \bibinfo{author}{L.~M.
  Lederman}, \bibinfo{author}{W.~Chinowsky}, \bibinfo{title}{{Long-lived
  neutral K mesons}}, \bibinfo{journal}{Annals Phys.}
  \bibinfo{volume}{5}~(\bibinfo{number}{2}) (\bibinfo{year}{1958})
  \bibinfo{pages}{156--181}.

\bibitem[{Lattes et~al.(1947)Lattes, Muirhead, Occhialini, and
  Powell}]{Lattes:1947mw}
\bibinfo{author}{C.~M.~G. Lattes}, \bibinfo{author}{H.~Muirhead},
  \bibinfo{author}{G.~P.~S. Occhialini}, \bibinfo{author}{C.~F. Powell},
  \bibinfo{title}{{Processes involving charged mesons}},
  \bibinfo{journal}{Nature} \bibinfo{volume}{159} (\bibinfo{year}{1947})
  \bibinfo{pages}{694--697}.

\bibitem[{Roberts et~al.(2021)Roberts, Richards, Horn, and
  Chang}]{Roberts:2021nhw}
\bibinfo{author}{C.~D. Roberts}, \bibinfo{author}{D.~G. Richards},
  \bibinfo{author}{T.~Horn}, \bibinfo{author}{L.~Chang},
  \bibinfo{title}{{Insights into the emergence of mass from studies of pion and
  kaon structure}}, \bibinfo{journal}{Prog. Part. Nucl. Phys.}
  \bibinfo{volume}{120} (\bibinfo{year}{2021}) \bibinfo{pages}{103883}.

\bibitem[{Binosi(2022)}]{Binosi:2022djx}
\bibinfo{author}{D.~Binosi}, \bibinfo{title}{{Emergent Hadron Mass in Strong
  Dynamics}}, \bibinfo{journal}{Few Body Syst.}
  \bibinfo{volume}{63}~(\bibinfo{number}{2}) (\bibinfo{year}{2022})
  \bibinfo{pages}{42}.

\bibitem[{Ding et~al.(2023)Ding, Roberts, and Schmidt}]{Ding:2022ows}
\bibinfo{author}{M.~Ding}, \bibinfo{author}{C.~D. Roberts},
  \bibinfo{author}{S.~M. Schmidt}, \bibinfo{title}{{Emergence of Hadron Mass
  and Structure}}, \bibinfo{journal}{Particles}
  \bibinfo{volume}{6}~(\bibinfo{number}{1}) (\bibinfo{year}{2023})
  \bibinfo{pages}{57--120}.

\bibitem[{Ferreira and Papavassiliou(2023)}]{Ferreira:2023fva}
\bibinfo{author}{M.~N. Ferreira}, \bibinfo{author}{J.~Papavassiliou},
  \bibinfo{title}{{Gauge Sector Dynamics in QCD}}, \bibinfo{journal}{Particles}
  \bibinfo{volume}{6}~(\bibinfo{number}{1}) (\bibinfo{year}{2023})
  \bibinfo{pages}{312--363}.

\bibitem[{Carman et~al.(2023)Carman, Gothe, Mokeev, and
  Roberts}]{Carman:2023zke}
\bibinfo{author}{D.~S. Carman}, \bibinfo{author}{R.~W. Gothe},
  \bibinfo{author}{V.~I. Mokeev}, \bibinfo{author}{C.~D. Roberts},
  \bibinfo{title}{{Nucleon Resonance Electroexcitation Amplitudes and Emergent
  Hadron Mass}}, \bibinfo{journal}{Particles}
  \bibinfo{volume}{6}~(\bibinfo{number}{1}) (\bibinfo{year}{2023})
  \bibinfo{pages}{416--439}.

\bibitem[{Raya et~al.(2024)Raya, Bashir, Binosi, Roberts, and
  Rodr\'\i{}guez-Quintero}]{Raya:2024ejx}
\bibinfo{author}{K.~Raya}, \bibinfo{author}{A.~Bashir},
  \bibinfo{author}{D.~Binosi}, \bibinfo{author}{C.~D. Roberts},
  \bibinfo{author}{J.~Rodr\'\i{}guez-Quintero}, \bibinfo{title}{{Pseudoscalar
  Mesons and Emergent Mass}}, \bibinfo{journal}{Few Body Syst.}
  \bibinfo{volume}{65}~(\bibinfo{number}{2}) (\bibinfo{year}{2024})
  \bibinfo{pages}{60}.

\bibitem[{Horn and Roberts(2016)}]{Horn:2016rip}
\bibinfo{author}{T.~Horn}, \bibinfo{author}{C.~D. Roberts},
  \bibinfo{title}{{The pion: an enigma within the Standard Model}},
  \bibinfo{journal}{J. Phys. G.} \bibinfo{volume}{43} (\bibinfo{year}{2016})
  \bibinfo{pages}{073001}.

\bibitem[{Cui et~al.(2021{\natexlab{a}})Cui, Binosi, Roberts, and
  Schmidt}]{Cui:2021aee}
\bibinfo{author}{Z.-F. Cui}, \bibinfo{author}{D.~Binosi},
  \bibinfo{author}{C.~D. Roberts}, \bibinfo{author}{S.~M. Schmidt},
  \bibinfo{title}{{Pion charge radius from pion+electron elastic scattering
  data}}, \bibinfo{journal}{Phys. Lett. B} \bibinfo{volume}{822}
  (\bibinfo{year}{2021}{\natexlab{a}}) \bibinfo{pages}{136631}.

\bibitem[{Carmignotto et~al.(2018)}]{Carmignotto:2018uqj}
\bibinfo{author}{M.~Carmignotto}, et~al., \bibinfo{title}{{Separated Kaon
  Electroproduction Cross Section and the Kaon Form Factor from 6 GeV JLab
  Data}}, \bibinfo{journal}{Phys. Rev. C} \bibinfo{volume}{97}
  (\bibinfo{year}{2018}) \bibinfo{pages}{025204}.

\bibitem[{Badier et~al.(1980)}]{Badier:1980jq}
\bibinfo{author}{J.~Badier}, et~al., \bibinfo{title}{{Measurement of the {$K^-
  / \pi^-$} structure function ratio using the Drell-Yan process}},
  \bibinfo{journal}{Phys. Lett. B} \bibinfo{volume}{93} (\bibinfo{year}{1980})
  \bibinfo{pages}{354}.

\bibitem[{{K. Park, R. Montgomery, T. Horn \emph{et al}.}(2015)}]{JlabTDIS2}
\bibinfo{author}{{K. Park, R. Montgomery, T. Horn \emph{et al}.}},
  \bibinfo{title}{Measurement of Kaon Structure Function through Tagged Deep
  Inelastic Scattering (\mbox{TDIS})} \bibinfo{note}{\mbox{ }approved Jefferson
  Lab experiment C12-15-006A.}

\bibitem[{Chen et~al.(2020)Chen, Guo, Roberts, and Wang}]{Chen:2020ijn}
\bibinfo{author}{X.~Chen}, \bibinfo{author}{F.-K. Guo}, \bibinfo{author}{C.~D.
  Roberts}, \bibinfo{author}{R.~Wang}, \bibinfo{title}{{Selected Science
  Opportunities for the EicC}}, \bibinfo{journal}{Few Body Syst.}
  \bibinfo{volume}{61} (\bibinfo{year}{2020}) \bibinfo{pages}{43}.

\bibitem[{Arrington et~al.(2021)}]{Arrington:2021biu}
\bibinfo{author}{J.~Arrington}, et~al., \bibinfo{title}{{Revealing the
  structure of light pseudoscalar mesons at the electron\textendash{}ion
  collider}}, \bibinfo{journal}{J. Phys. G} \bibinfo{volume}{48}
  (\bibinfo{year}{2021}) \bibinfo{pages}{075106}.

\bibitem[{Seitz(2023)}]{Seitz:2023rqm}
\bibinfo{author}{B.~Seitz}, \bibinfo{title}{{AMBER: a new QCD facility at the
  CERN SPS M2 beam line}}, \bibinfo{journal}{PoS} \bibinfo{volume}{ICHEP2022}
  (\bibinfo{year}{2023}) \bibinfo{pages}{839}.

\bibitem[{Metzger et~al.(2023)}]{Metzger:2023vxl}
\bibinfo{author}{F.~Metzger}, et~al., \bibinfo{title}{{Kaon beam simulations
  employing conventional hadron beam concepts and the RF separation technique
  at the CERN M2 beamline for the future AMBER experiment}},
  \bibinfo{journal}{JACoW} \bibinfo{volume}{IPAC2023} (\bibinfo{year}{2023})
  \bibinfo{pages}{TUPM070}.

\bibitem[{Kock et~al.(2020)Kock, Liu, and Zahed}]{Kock:2020frx}
\bibinfo{author}{A.~Kock}, \bibinfo{author}{Y.~Liu},
  \bibinfo{author}{I.~Zahed}, \bibinfo{title}{{Pion and kaon parton
  distributions in the QCD instanton vacuum}}, \bibinfo{journal}{Phys. Rev. D}
  \bibinfo{volume}{102}~(\bibinfo{number}{1}) (\bibinfo{year}{2020})
  \bibinfo{pages}{014039}.

\bibitem[{Cui et~al.(2020{\natexlab{a}})Cui, Ding, Gao, Raya, Binosi, Chang,
  Roberts, Rodr\'{\i}guez-Quintero, and Schmidt}]{Cui:2020tdf}
\bibinfo{author}{Z.-F. Cui}, \bibinfo{author}{M.~Ding},
  \bibinfo{author}{F.~Gao}, \bibinfo{author}{K.~Raya},
  \bibinfo{author}{D.~Binosi}, \bibinfo{author}{L.~Chang},
  \bibinfo{author}{C.~D. Roberts},
  \bibinfo{author}{J.~Rodr\'{\i}guez-Quintero}, \bibinfo{author}{S.~M.
  Schmidt}, \bibinfo{title}{{Kaon and pion parton distributions}},
  \bibinfo{journal}{Eur. Phys. J. C} \bibinfo{volume}{80}
  (\bibinfo{year}{2020}{\natexlab{a}}) \bibinfo{pages}{1064}.

\bibitem[{Cui et~al.(2021{\natexlab{b}})Cui, Ding, Gao, Raya, Binosi, Chang,
  Roberts, Rodr\'{\i}guez-Quintero, and Schmidt}]{Cui:2020dlm}
\bibinfo{author}{Z.-F. Cui}, \bibinfo{author}{M.~Ding},
  \bibinfo{author}{F.~Gao}, \bibinfo{author}{K.~Raya},
  \bibinfo{author}{D.~Binosi}, \bibinfo{author}{L.~Chang},
  \bibinfo{author}{C.~D. Roberts},
  \bibinfo{author}{J.~Rodr\'{\i}guez-Quintero}, \bibinfo{author}{S.~M.
  Schmidt}, \bibinfo{title}{{Higgs modulation of emergent mass as revealed in
  kaon and pion parton distributions}}, \bibinfo{journal}{Eur. Phys. J. A
  (Lett.)} \bibinfo{volume}{57}~(\bibinfo{number}{1})
  (\bibinfo{year}{2021}{\natexlab{b}}) \bibinfo{pages}{5}.

\bibitem[{Lin et~al.(2021)Lin, Chen, Fan, Zhang, and Zhang}]{Lin:2020ssv}
\bibinfo{author}{H.-W. Lin}, \bibinfo{author}{J.-W. Chen},
  \bibinfo{author}{Z.~Fan}, \bibinfo{author}{J.-H. Zhang},
  \bibinfo{author}{R.~Zhang}, \bibinfo{title}{{Valence-Quark Distribution of
  the Kaon and Pion from Lattice QCD}}, \bibinfo{journal}{Phys. Rev. D}
  \bibinfo{volume}{103}~(\bibinfo{number}{1}) (\bibinfo{year}{2021})
  \bibinfo{pages}{014516}.

\bibitem[{Xie et~al.(2022)Xie, Han, Wang, and Chen}]{Xie:2021ypc}
\bibinfo{author}{G.~Xie}, \bibinfo{author}{C.~Han}, \bibinfo{author}{R.~Wang},
  \bibinfo{author}{X.~Chen}, \bibinfo{title}{{Tackling the kaon structure
  function at EicC}}, \bibinfo{journal}{Chin. Phys. C}
  \bibinfo{volume}{46}~(\bibinfo{number}{6}) (\bibinfo{year}{2022})
  \bibinfo{pages}{064107}.

\bibitem[{Ch\'avez et~al.(2022)Ch\'avez, Bertone, De~Soto~Borrero, Defurne,
  Mezrag, Moutarde, Rodr\'\i{}guez-Quintero, and Segovia}]{Chavez:2021koz}
\bibinfo{author}{J.~M.~M. Ch\'avez}, \bibinfo{author}{V.~Bertone},
  \bibinfo{author}{F.~De~Soto~Borrero}, \bibinfo{author}{M.~Defurne},
  \bibinfo{author}{C.~Mezrag}, \bibinfo{author}{H.~Moutarde},
  \bibinfo{author}{J.~Rodr\'\i{}guez-Quintero}, \bibinfo{author}{J.~Segovia},
  \bibinfo{title}{{Accessing the Pion 3D Structure at US and China Electron-Ion
  Colliders}}, \bibinfo{journal}{Phys. Rev. Lett.}
  \bibinfo{volume}{128}~(\bibinfo{number}{20}) (\bibinfo{year}{2022})
  \bibinfo{pages}{202501}.

\bibitem[{de~Paula et~al.(2022)de~Paula, Ydrefors, Nogueira~Alvarenga,
  Frederico, and Salm\`e}]{dePaula:2022pcb}
\bibinfo{author}{W.~de~Paula}, \bibinfo{author}{E.~Ydrefors},
  \bibinfo{author}{J.~H. Nogueira~Alvarenga}, \bibinfo{author}{T.~Frederico},
  \bibinfo{author}{G.~Salm\`e}, \bibinfo{title}{{Parton distribution function
  in a pion with Minkowskian dynamics}}, \bibinfo{journal}{Phys. Rev. D}
  \bibinfo{volume}{105}~(\bibinfo{number}{7}) (\bibinfo{year}{2022})
  \bibinfo{pages}{L071505}.

\bibitem[{Pasquini et~al.(2023)Pasquini, Rodini, and
  Venturini}]{Pasquini:2023aaf}
\bibinfo{author}{B.~Pasquini}, \bibinfo{author}{S.~Rodini},
  \bibinfo{author}{S.~Venturini}, \bibinfo{title}{{Valence quark, sea, and
  gluon content of the pion from the parton distribution functions and the
  electromagnetic form factor}}, \bibinfo{journal}{Phys. Rev. D}
  \bibinfo{volume}{107}~(\bibinfo{number}{11}) (\bibinfo{year}{2023})
  \bibinfo{pages}{114023}.

\bibitem[{Wang et~al.(2024)Wang, Ding, and Chang}]{Wang:2023bmk}
\bibinfo{author}{X.~Wang}, \bibinfo{author}{M.~Ding},
  \bibinfo{author}{L.~Chang}, \bibinfo{title}{{Sieving parton distribution
  function moments via the moment problem}}, \bibinfo{journal}{Phys. Lett. B}
  \bibinfo{volume}{851} (\bibinfo{year}{2024}) \bibinfo{pages}{138568}.

\bibitem[{Good et~al.(2024)Good, Hasan, Chevis, and Lin}]{Good:2023ecp}
\bibinfo{author}{W.~Good}, \bibinfo{author}{K.~Hasan},
  \bibinfo{author}{A.~Chevis}, \bibinfo{author}{H.-W. Lin},
  \bibinfo{title}{{Gluon moment and parton distribution function of the pion
  from Nf=2+1+1 lattice QCD}}, \bibinfo{journal}{Phys. Rev. D}
  \bibinfo{volume}{109}~(\bibinfo{number}{11}) (\bibinfo{year}{2024})
  \bibinfo{pages}{114509}.

\bibitem[{Chang et~al.(2024)Chang, Peng, Platchkov, and Sawada}]{Chang:2024rbs}
\bibinfo{author}{W.-C. Chang}, \bibinfo{author}{J.-C. Peng},
  \bibinfo{author}{S.~Platchkov}, \bibinfo{author}{T.~Sawada},
  \bibinfo{title}{{Constraining kaon PDFs from Drell-Yan and
  J/\ensuremath{\psi} production}}, \bibinfo{journal}{Phys. Lett. B}
  \bibinfo{volume}{855} (\bibinfo{year}{2024}) \bibinfo{pages}{138820}.

\bibitem[{Alexandrou et~al.(2024)}]{Alexandrou:2024zvn}
\bibinfo{author}{C.~Alexandrou}, et~al., \bibinfo{title}{{Quark and gluon
  momentum fractions in the pion and in the kaon -- arXiv:2405.08529
  [hep-lat]}} .

\bibitem[{Dokshitzer(1977)}]{Dokshitzer:1977sg}
\bibinfo{author}{Y.~L. Dokshitzer}, \bibinfo{title}{Calculation of the
  Structure Functions for Deep Inelastic Scattering and $e^+$ $e^-$
  Annihilation by Perturbation Theory in Quantum Chromodynamics. ({\mbox {I}n
  {R}ussian})}, \bibinfo{journal}{Sov. Phys. JETP} \bibinfo{volume}{46}
  (\bibinfo{year}{1977}) \bibinfo{pages}{641--653}.

\bibitem[{Gribov and Lipatov(1971)}]{Gribov:1971zn}
\bibinfo{author}{V.~N. Gribov}, \bibinfo{author}{L.~N. Lipatov},
  \bibinfo{title}{{Deep inelastic electron scattering in perturbation theory}},
  \bibinfo{journal}{Phys. Lett. B} \bibinfo{volume}{37} (\bibinfo{year}{1971})
  \bibinfo{pages}{78--80}.

\bibitem[{Lipatov(1975)}]{Lipatov:1974qm}
\bibinfo{author}{L.~N. Lipatov}, \bibinfo{title}{{The parton model and
  perturbation theory}}, \bibinfo{journal}{Sov. J. Nucl. Phys.}
  \bibinfo{volume}{20} (\bibinfo{year}{1975}) \bibinfo{pages}{94--102}.

\bibitem[{Altarelli and Parisi(1977)}]{Altarelli:1977zs}
\bibinfo{author}{G.~Altarelli}, \bibinfo{author}{G.~Parisi},
  \bibinfo{title}{{Asymptotic Freedom in Parton Language}},
  \bibinfo{journal}{Nucl. Phys. B} \bibinfo{volume}{126} (\bibinfo{year}{1977})
  \bibinfo{pages}{298--318}.

\bibitem[{Yin et~al.(2023)Yin, Xu, Cui, Roberts, and
  Rodr\'\i{}guez-Quintero}]{Yin:2023dbw}
\bibinfo{author}{P.-L. Yin}, \bibinfo{author}{Y.-Z. Xu}, \bibinfo{author}{Z.-F.
  Cui}, \bibinfo{author}{C.~D. Roberts},
  \bibinfo{author}{J.~Rodr\'\i{}guez-Quintero}, \bibinfo{title}{{All-Orders
  Evolution of Parton Distributions: Principle, Practice, and Predictions}},
  \bibinfo{journal}{Chin. Phys. Lett. \emph{Express}}
  \bibinfo{volume}{40}~(\bibinfo{number}{9}) (\bibinfo{year}{2023})
  \bibinfo{pages}{091201}.

\bibitem[{Grunberg(1980)}]{Grunberg:1980ja}
\bibinfo{author}{G.~Grunberg}, \bibinfo{title}{{Renormalization Group Improved
  Perturbative QCD}}, \bibinfo{journal}{Phys. Lett. B} \bibinfo{volume}{95}
  (\bibinfo{year}{1980}) \bibinfo{pages}{70}, \bibinfo{note}{[Erratum: Phys.
  Lett. B 110, 501 (1982)]}.

\bibitem[{Grunberg(1984)}]{Grunberg:1982fw}
\bibinfo{author}{G.~Grunberg}, \bibinfo{title}{{Renormalization Scheme
  Independent QCD and QED: The Method of Effective Charges}},
  \bibinfo{journal}{Phys. Rev. D} \bibinfo{volume}{29} (\bibinfo{year}{1984})
  \bibinfo{pages}{2315}.

\bibitem[{Deur et~al.(2024)Deur, Brodsky, and Roberts}]{Deur:2023dzc}
\bibinfo{author}{A.~Deur}, \bibinfo{author}{S.~J. Brodsky},
  \bibinfo{author}{C.~D. Roberts}, \bibinfo{title}{{QCD Running Couplings and
  Effective Charges}}, \bibinfo{journal}{Prog. Part. Nucl. Phys.}
  \bibinfo{volume}{134} (\bibinfo{year}{2024}) \bibinfo{pages}{104081}.

\bibitem[{Cui et~al.(2020{\natexlab{b}})Cui, Zhang, Binosi, de~Soto, Mezrag,
  Papavassiliou, Roberts, Rodr{\'{\i}}guez-Quintero, Segovia, and
  Zafeiropoulos}]{Cui:2019dwv}
\bibinfo{author}{Z.-F. Cui}, \bibinfo{author}{J.-L. Zhang},
  \bibinfo{author}{D.~Binosi}, \bibinfo{author}{F.~de~Soto},
  \bibinfo{author}{C.~Mezrag}, \bibinfo{author}{J.~Papavassiliou},
  \bibinfo{author}{C.~D. Roberts},
  \bibinfo{author}{J.~Rodr{\'{\i}}guez-Quintero}, \bibinfo{author}{J.~Segovia},
  \bibinfo{author}{S.~Zafeiropoulos}, \bibinfo{title}{{Effective charge from
  lattice QCD}}, \bibinfo{journal}{Chin. Phys. C} \bibinfo{volume}{44}
  (\bibinfo{year}{2020}{\natexlab{b}}) \bibinfo{pages}{083102}.

\bibitem[{Brodsky et~al.(2024)Brodsky, Deur, and Roberts}]{Brodsky:2024zev}
\bibinfo{author}{S.~J. Brodsky}, \bibinfo{author}{A.~Deur},
  \bibinfo{author}{C.~D. Roberts}, \bibinfo{title}{{The Secret to the Strongest
  Force in the Universe}}, \bibinfo{journal}{Sci. Am.} \bibinfo{volume}{5
  (May)} (\bibinfo{year}{2024}) \bibinfo{pages}{32--39}.

\bibitem[{Cui et~al.(2022{\natexlab{a}})Cui, Ding, Morgado, Raya, Binosi,
  Chang, Papavassiliou, Roberts, Rodr\'\i{}guez-Quintero, and
  Schmidt}]{Cui:2021mom}
\bibinfo{author}{Z.~F. Cui}, \bibinfo{author}{M.~Ding}, \bibinfo{author}{J.~M.
  Morgado}, \bibinfo{author}{K.~Raya}, \bibinfo{author}{D.~Binosi},
  \bibinfo{author}{L.~Chang}, \bibinfo{author}{J.~Papavassiliou},
  \bibinfo{author}{C.~D. Roberts},
  \bibinfo{author}{J.~Rodr\'\i{}guez-Quintero}, \bibinfo{author}{S.~M.
  Schmidt}, \bibinfo{title}{{Concerning pion parton distributions}},
  \bibinfo{journal}{Eur. Phys. J. A} \bibinfo{volume}{58}~(\bibinfo{number}{1})
  (\bibinfo{year}{2022}{\natexlab{a}}) \bibinfo{pages}{10}.

\bibitem[{Lu et~al.(2024)Lu, Xu, Raya, Roberts, and
  Rodr\'\i{}guez-Quintero}]{Lu:2023yna}
\bibinfo{author}{Y.~Lu}, \bibinfo{author}{Y.-Z. Xu}, \bibinfo{author}{K.~Raya},
  \bibinfo{author}{C.~D. Roberts},
  \bibinfo{author}{J.~Rodr\'\i{}guez-Quintero}, \bibinfo{title}{{Pion
  distribution functions from low-order Mellin moments}},
  \bibinfo{journal}{Phys. Lett. B} \bibinfo{volume}{850} (\bibinfo{year}{2024})
  \bibinfo{pages}{138534}.

\bibitem[{Xing et~al.(2024)Xing, Yao, Li, Binosi, Cui, and
  Roberts}]{Xing:2023pms}
\bibinfo{author}{H.~Y. Xing}, \bibinfo{author}{Z.~Q. Yao},
  \bibinfo{author}{B.~L. Li}, \bibinfo{author}{D.~Binosi},
  \bibinfo{author}{Z.~F. Cui}, \bibinfo{author}{C.~D. Roberts},
  \bibinfo{title}{{Developing predictions for pion fragmentation functions}},
  \bibinfo{journal}{Eur. Phys. J. C} \bibinfo{volume}{84}~(\bibinfo{number}{1})
  (\bibinfo{year}{2024}) \bibinfo{pages}{82}.

\bibitem[{Chang et~al.(2022)Chang, Gao, and Roberts}]{Chang:2022jri}
\bibinfo{author}{L.~Chang}, \bibinfo{author}{F.~Gao}, \bibinfo{author}{C.~D.
  Roberts}, \bibinfo{title}{{Parton distributions of light quarks and
  antiquarks in the proton}}, \bibinfo{journal}{Phys. Lett. B}
  \bibinfo{volume}{829} (\bibinfo{year}{2022}) \bibinfo{pages}{137078}.

\bibitem[{Lu et~al.(2022)Lu, Chang, Raya, Roberts, and
  Rodr\'\i{}guez-Quintero}]{Lu:2022cjx}
\bibinfo{author}{Y.~Lu}, \bibinfo{author}{L.~Chang}, \bibinfo{author}{K.~Raya},
  \bibinfo{author}{C.~D. Roberts},
  \bibinfo{author}{J.~Rodr\'\i{}guez-Quintero}, \bibinfo{title}{{Proton and
  pion distribution functions in counterpoint}}, \bibinfo{journal}{Phys. Lett.
  B} \bibinfo{volume}{830} (\bibinfo{year}{2022}) \bibinfo{pages}{137130}.

\bibitem[{Cheng et~al.(2023)Cheng, Yu, Xing, Chen, Cui, and
  Roberts}]{Cheng:2023kmt}
\bibinfo{author}{P.~Cheng}, \bibinfo{author}{Y.~Yu}, \bibinfo{author}{H.-Y.
  Xing}, \bibinfo{author}{C.~Chen}, \bibinfo{author}{Z.-F. Cui},
  \bibinfo{author}{C.~D. Roberts}, \bibinfo{title}{{Perspective on polarised
  parton distribution functions and proton spin}}, \bibinfo{journal}{Phys.
  Lett. B} \bibinfo{volume}{844} (\bibinfo{year}{2023})
  \bibinfo{pages}{138074}.

\bibitem[{Yu et~al.(2024)Yu, Cheng, Xing, Gao, and Roberts}]{Yu:2024qsd}
\bibinfo{author}{Y.~Yu}, \bibinfo{author}{P.~Cheng}, \bibinfo{author}{H.-Y.
  Xing}, \bibinfo{author}{F.~Gao}, \bibinfo{author}{C.~D. Roberts},
  \bibinfo{title}{{Contact interaction study of proton parton distributions}},
  \bibinfo{journal}{Eur. Phys. J. C} \bibinfo{volume}{84}~(\bibinfo{number}{7})
  (\bibinfo{year}{2024}) \bibinfo{pages}{739}.

\bibitem[{Yao et~al.(2024)Yao, Xu, Binosi, Cui, Ding, Raya, Roberts,
  Rodr\'\i{}guez-Quintero, and Schmidt}]{Yao:2024ixu}
\bibinfo{author}{Z.~Q. Yao}, \bibinfo{author}{Y.~Z. Xu},
  \bibinfo{author}{D.~Binosi}, \bibinfo{author}{Z.~F. Cui},
  \bibinfo{author}{M.~Ding}, \bibinfo{author}{K.~Raya}, \bibinfo{author}{C.~D.
  Roberts}, \bibinfo{author}{J.~Rodr\'\i{}guez-Quintero},
  \bibinfo{author}{S.~M. Schmidt}, \bibinfo{title}{{Nucleon Gravitational Form
  Factors -- arXiv:2409.15547 [hep-ph]}} .

\bibitem[{Yu and Roberts(2024)}]{Yu:2024ovn}
\bibinfo{author}{Y.~Yu}, \bibinfo{author}{C.~D. Roberts},
  \bibinfo{title}{{Impressions of Parton Distribution Functions --
  arXiv:2410.03966 [hep-ph]}}, \bibinfo{journal}{Chin. Phys. Lett.}
  (\bibinfo{year}{2024}) \bibinfo{pages}{\emph{in press}}.

\bibitem[{Cui et~al.(2022{\natexlab{b}})Cui, Ding, Morgado, Raya, Binosi,
  Chang, De~Soto, Roberts, Rodr\'\i{}guez-Quintero, and Schmidt}]{Cui:2022bxn}
\bibinfo{author}{Z.~F. Cui}, \bibinfo{author}{M.~Ding}, \bibinfo{author}{J.~M.
  Morgado}, \bibinfo{author}{K.~Raya}, \bibinfo{author}{D.~Binosi},
  \bibinfo{author}{L.~Chang}, \bibinfo{author}{F.~De~Soto},
  \bibinfo{author}{C.~D. Roberts},
  \bibinfo{author}{J.~Rodr\'\i{}guez-Quintero}, \bibinfo{author}{S.~M.
  Schmidt}, \bibinfo{title}{{Emergence of pion parton distributions}},
  \bibinfo{journal}{Phys. Rev. D} \bibinfo{volume}{105}~(\bibinfo{number}{9})
  (\bibinfo{year}{2022}{\natexlab{b}}) \bibinfo{pages}{L091502}.

\bibitem[{Conway et~al.(1989)}]{Conway:1989fs}
\bibinfo{author}{J.~S. Conway}, et~al., \bibinfo{title}{{Experimental study of
  muon pairs produced by 252-GeV pions on tungsten}}, \bibinfo{journal}{Phys.
  Rev. D} \bibinfo{volume}{39} (\bibinfo{year}{1989}) \bibinfo{pages}{92--122}.

\bibitem[{Aicher et~al.(2010)Aicher, Sch{\"a}fer, and
  Vogelsang}]{Aicher:2010cb}
\bibinfo{author}{M.~Aicher}, \bibinfo{author}{A.~Sch{\"a}fer},
  \bibinfo{author}{W.~Vogelsang}, \bibinfo{title}{{Soft-Gluon Resummation and
  the Valence Parton Distribution Function of the Pion}},
  \bibinfo{journal}{Phys.\ Rev.\ Lett.} \bibinfo{volume}{105}
  (\bibinfo{year}{2010}) \bibinfo{pages}{252003}.

\bibitem[{Chang et~al.(2014)Chang, Mezrag, Moutarde, Roberts,
  Rodr{\'{\i}}guez-Quintero, and Tandy}]{Chang:2014lva}
\bibinfo{author}{L.~Chang}, \bibinfo{author}{C.~Mezrag},
  \bibinfo{author}{H.~Moutarde}, \bibinfo{author}{C.~D. Roberts},
  \bibinfo{author}{J.~Rodr{\'{\i}}guez-Quintero}, \bibinfo{author}{P.~C.
  Tandy}, \bibinfo{title}{{Basic features of the pion valence-quark
  distribution function}}, \bibinfo{journal}{Phys. Lett. B}
  \bibinfo{volume}{737} (\bibinfo{year}{2014}) \bibinfo{pages}{23--29}.

\bibitem[{Brodsky et~al.(1995)Brodsky, Burkardt, and Schmidt}]{Brodsky:1994kg}
\bibinfo{author}{S.~J. Brodsky}, \bibinfo{author}{M.~Burkardt},
  \bibinfo{author}{I.~Schmidt}, \bibinfo{title}{{Perturbative QCD constraints
  on the shape of polarized quark and gluon distributions}},
  \bibinfo{journal}{Nucl. Phys. B} \bibinfo{volume}{441} (\bibinfo{year}{1995})
  \bibinfo{pages}{197--214}.

\bibitem[{Yuan(2004)}]{Yuan:2003fs}
\bibinfo{author}{F.~Yuan}, \bibinfo{title}{{Generalized parton distributions at
  $x \to 1$}}, \bibinfo{journal}{Phys. Rev. D} \bibinfo{volume}{69}
  (\bibinfo{year}{2004}) \bibinfo{pages}{051501}.

\bibitem[{Holt and Roberts(2010)}]{Holt:2010vj}
\bibinfo{author}{R.~J. Holt}, \bibinfo{author}{C.~D. Roberts},
  \bibinfo{title}{{Distribution Functions of the Nucleon and Pion in the
  Valence Region}}, \bibinfo{journal}{Rev. Mod. Phys.} \bibinfo{volume}{82}
  (\bibinfo{year}{2010}) \bibinfo{pages}{2991--3044}.

\bibitem[{Jo\'o et~al.(2019)Jo\'o, Karpie, Orginos, Radyushkin, Richards,
  Sufian, and Zafeiropoulos}]{Joo:2019bzr}
\bibinfo{author}{B.~Jo\'o}, \bibinfo{author}{J.~Karpie},
  \bibinfo{author}{K.~Orginos}, \bibinfo{author}{A.~V. Radyushkin},
  \bibinfo{author}{D.~G. Richards}, \bibinfo{author}{R.~S. Sufian},
  \bibinfo{author}{S.~Zafeiropoulos}, \bibinfo{title}{{Pion valence structure
  from Ioffe-time parton pseudodistribution functions}},
  \bibinfo{journal}{Phys. Rev. D} \bibinfo{volume}{100} (\bibinfo{year}{2019})
  \bibinfo{pages}{114512}.

\bibitem[{Sufian et~al.(2019)Sufian, Karpie, Egerer, Orginos, Qiu, and
  Richards}]{Sufian:2019bol}
\bibinfo{author}{R.~S. Sufian}, \bibinfo{author}{J.~Karpie},
  \bibinfo{author}{C.~Egerer}, \bibinfo{author}{K.~Orginos},
  \bibinfo{author}{J.-W. Qiu}, \bibinfo{author}{D.~G. Richards},
  \bibinfo{title}{{Pion Valence Quark Distribution from Matrix Element
  Calculated in Lattice QCD}}, \bibinfo{journal}{Phys. Rev. D}
  \bibinfo{volume}{99} (\bibinfo{year}{2019}) \bibinfo{pages}{074507}.

\bibitem[{Alexandrou et~al.(2021{\natexlab{a}})Alexandrou, Bacchio, Cloet,
  Constantinou, Hadjiyiannakou, Koutsou, and Lauer}]{Alexandrou:2021mmi}
\bibinfo{author}{C.~Alexandrou}, \bibinfo{author}{S.~Bacchio},
  \bibinfo{author}{I.~Cloet}, \bibinfo{author}{M.~Constantinou},
  \bibinfo{author}{K.~Hadjiyiannakou}, \bibinfo{author}{G.~Koutsou},
  \bibinfo{author}{C.~Lauer}, \bibinfo{title}{{Pion and kaon $\langle
  x^3\rangle$ from lattice QCD and PDF reconstruction from Mellin moments}},
  \bibinfo{journal}{Phys. Rev. D} \bibinfo{volume}{104}~(\bibinfo{number}{5})
  (\bibinfo{year}{2021}{\natexlab{a}}) \bibinfo{pages}{054504}.

\bibitem[{Gao et~al.(2022)Gao, Hanlon, Karthik, Mukherjee, Petreczky, Scior,
  Shi, Syritsyn, Zhao, and Zhou}]{Gao:2022iex}
\bibinfo{author}{X.~Gao}, \bibinfo{author}{A.~D. Hanlon},
  \bibinfo{author}{N.~Karthik}, \bibinfo{author}{S.~Mukherjee},
  \bibinfo{author}{P.~Petreczky}, \bibinfo{author}{P.~Scior},
  \bibinfo{author}{S.~Shi}, \bibinfo{author}{S.~Syritsyn},
  \bibinfo{author}{Y.~Zhao}, \bibinfo{author}{K.~Zhou},
  \bibinfo{title}{{Continuum-extrapolated NNLO valence PDF of the pion at the
  physical point}}, \bibinfo{journal}{Phys. Rev. D}
  \bibinfo{volume}{106}~(\bibinfo{number}{11}) (\bibinfo{year}{2022})
  \bibinfo{pages}{114510}.

\bibitem[{Lepage and Brodsky(1979)}]{Lepage:1979zb}
\bibinfo{author}{G.~P. Lepage}, \bibinfo{author}{S.~J. Brodsky},
  \bibinfo{title}{{Exclusive Processes in Quantum Chromodynamics: Evolution
  Equations for Hadronic Wave Functions and the Form-Factors of Mesons}},
  \bibinfo{journal}{Phys. Lett. B} \bibinfo{volume}{87} (\bibinfo{year}{1979})
  \bibinfo{pages}{359--365}.

\bibitem[{Efremov and Radyushkin(1980)}]{Efremov:1979qk}
\bibinfo{author}{A.~V. Efremov}, \bibinfo{author}{A.~V. Radyushkin},
  \bibinfo{title}{{Factorization and Asymptotical Behavior of Pion Form- Factor
  in QCD}}, \bibinfo{journal}{Phys. Lett. B} \bibinfo{volume}{94}
  (\bibinfo{year}{1980}) \bibinfo{pages}{245--250}.

\bibitem[{Lepage and Brodsky(1980)}]{Lepage:1980fj}
\bibinfo{author}{G.~P. Lepage}, \bibinfo{author}{S.~J. Brodsky},
  \bibinfo{title}{{Exclusive Processes in Perturbative Quantum
  Chromodynamics}}, \bibinfo{journal}{Phys. Rev. D} \bibinfo{volume}{22}
  (\bibinfo{year}{1980}) \bibinfo{pages}{2157--2198}.

\bibitem[{Navas et~al.(2024)}]{ParticleDataGroup:2024cfk}
\bibinfo{author}{S.~Navas}, et~al., \bibinfo{title}{{Review of particle
  physics}}, \bibinfo{journal}{Phys. Rev. D}
  \bibinfo{volume}{110}~(\bibinfo{number}{3}) (\bibinfo{year}{2024})
  \bibinfo{pages}{030001}.

\bibitem[{Alexandrou et~al.(2021{\natexlab{b}})Alexandrou, Bacchio, Cloet,
  Constantinou, Hadjiyiannakou, Koutsou, and Lauer}]{Alexandrou:2020gxs}
\bibinfo{author}{C.~Alexandrou}, \bibinfo{author}{S.~Bacchio},
  \bibinfo{author}{I.~Cloet}, \bibinfo{author}{M.~Constantinou},
  \bibinfo{author}{K.~Hadjiyiannakou}, \bibinfo{author}{G.~Koutsou},
  \bibinfo{author}{C.~Lauer}, \bibinfo{title}{{Mellin moments $\langle x
  \rangle$ and $\langle x^2 \rangle$ for the pion and kaon from lattice QCD}},
  \bibinfo{journal}{Phys. Rev. D} \bibinfo{volume}{103}~(\bibinfo{number}{1})
  (\bibinfo{year}{2021}{\natexlab{b}}) \bibinfo{pages}{014508}.

\bibitem[{Alexandrou(2024)}]{AlexandrouBNL}
\bibinfo{author}{C.~Alexandrou}, \bibinfo{title}{Presentation at the CFNS
  Workshop: Elucidating the Structure of Nambu-Goldstone Bosons --
  \href{https://indico.cfnssbu.physics.sunysb.edu/event/251/contributions/1022/attachments/378/618/BNL.pdf}
  {Mon.\ 2024/06/24}} .

\bibitem[{Chen et~al.(2016)Chen, Chang, Roberts, Wan, and Zong}]{Chen:2016sno}
\bibinfo{author}{C.~Chen}, \bibinfo{author}{L.~Chang}, \bibinfo{author}{C.~D.
  Roberts}, \bibinfo{author}{S.~Wan}, \bibinfo{author}{H.-S. Zong},
  \bibinfo{title}{{Valence-quark distribution functions in the kaon and pion}},
  \bibinfo{journal}{Phys. Rev. D} \bibinfo{volume}{93} (\bibinfo{year}{2016})
  \bibinfo{pages}{074021}.

\bibitem[{Gl{\"u}ck et~al.(1999)Gl{\"u}ck, Reya, and Schienbein}]{Gluck:1999xe}
\bibinfo{author}{M.~Gl{\"u}ck}, \bibinfo{author}{E.~Reya},
  \bibinfo{author}{I.~Schienbein}, \bibinfo{title}{{Pionic parton distributions
  revisited}}, \bibinfo{journal}{Eur. Phys. J. C} \bibinfo{volume}{10}
  (\bibinfo{year}{1999}) \bibinfo{pages}{313--317}.

\end{thebibliography}

\end{document}